\documentclass[manuscript,screen]{acmart}
\usepackage{algorithm}
\usepackage{algorithmic}
\usepackage{multirow}
\usepackage{tcolorbox}   
\usepackage{xcolor}  
\usepackage{ulem}
\AtBeginDocument{%
  }

\setcopyright{acmlicensed}
\copyrightyear{2025}
\acmYear{2025}
\acmDOI{XXXXXXX.XXXXXXX}
\acmISBN{978-1-4503-XXXX-X/2018/06}

\begin{document}

\title{FIRE: Faithful Interpretable Recommendation Explanations}

\author{S.M.F. Sani}

\email{s.feyzabadisani76@sharif.edu}
\orcid{0009-0000-5833-0270}
\affiliation{%
  \institution{Sharif University of Technology}
  \city{Tehran}
  \state{Tehran}
  \country{Iran}
}

\author{Asal Meskin}
\email{asal.meskin82@sharif.edu}
\affiliation{%
  \institution{Sharif University of Technology}
  \city{Tehran}
  \state{Tehran}
  \country{Iran}
}
\authornote{Both authors contributed equally to this research.}
\author{Mohammad Amanlou}
\email{Mohammad.amanlou@ut.ac.ir}
\authornotemark[1]
\affiliation{%
  \institution{University of Tehran}
  \city{Tehran}
  \state{Tehran}
  \country{Iran}
}

\author{Hamid. R. Rabiee}
\affiliation{%
  \institution{Sharif University of Technology}
  \city{Tehran}
  \state{Tehran}
  \country{Iran}
}
\email{rabiee@sharif.edu}


\begin{abstract}
  Natural language explanations in recommender systems are often framed as a review generation task, leveraging user reviews as ground-truth supervision. While convenient, this approach conflates a user’s opinion with the system’s reasoning, leading to explanations that may be fluent but fail to reflect the true logic behind recommendations. In this work, we revisit the core objective of explainable recommendation: to transparently communicate why an item is recommended by linking user needs to relevant item features. Through a comprehensive analysis of existing methods across multiple benchmark datasets, we identify common limitations—explanations that are weakly aligned with model predictions, vague or inaccurate in identifying user intents, and overly repetitive or generic. To overcome these challenges, we propose FIRE, a lightweight and interpretable framework that combines SHAP-based feature attribution with structured, prompt-driven language generation. FIRE produces faithful, diverse, and user-aligned explanations, grounded in the actual decision-making process of the model. Our results demonstrate that FIRE not only achieves competitive recommendation accuracy but also significantly improves explanation quality along critical dimensions such as alignment, structure, and faithfulness. This work highlights the need to move beyond the review-as-explanation paradigm and toward explanation methods that are both accountable and interpretable.
\end{abstract}

\ccsdesc[500]{Information systems~Recommender systems}
\ccsdesc[500]{Computing methodologies~Natural language generation}

\keywords{Explainable Recommender Systems}

\received{7 August 2025}

\maketitle

\section{Introduction} \label{sec:introduction}

Recommender systems play a crucial role in guiding users through vast information spaces by suggesting relevant products, services, or content. As these systems increasingly influence user behavior and decision-making, the demand for transparency and trust has led to growing interest in explainable recommendation, particularly through natural language explanations~\cite{li2021personalized, li2023personalized, xie2023factual, cheng2023explainable, geng2022recommendation}.

A widely adopted paradigm in recent literature—especially among approaches leveraging Large Language Models (LLMs)—treats review generation as a proxy for explanation. This assumes that a review inherently encapsulates the rationale behind user preferences and thus serves as a surrogate for explainability.

While this strategy has enabled rapid progress, offering a practical supervised training setup and the ability to mimic natural language fluently, it deviates from the core goal of explainability. Generating text in the style of a review does not ensure that the explanation faithfully reflects the model’s internal decision-making. Reviews are subjective, experiential, and often emotionally driven, whereas explanations should highlight the specific user-item interactions and features that influenced the recommendation.

\begin{figure}
    \centering
    \includegraphics[width=0.8\textwidth]{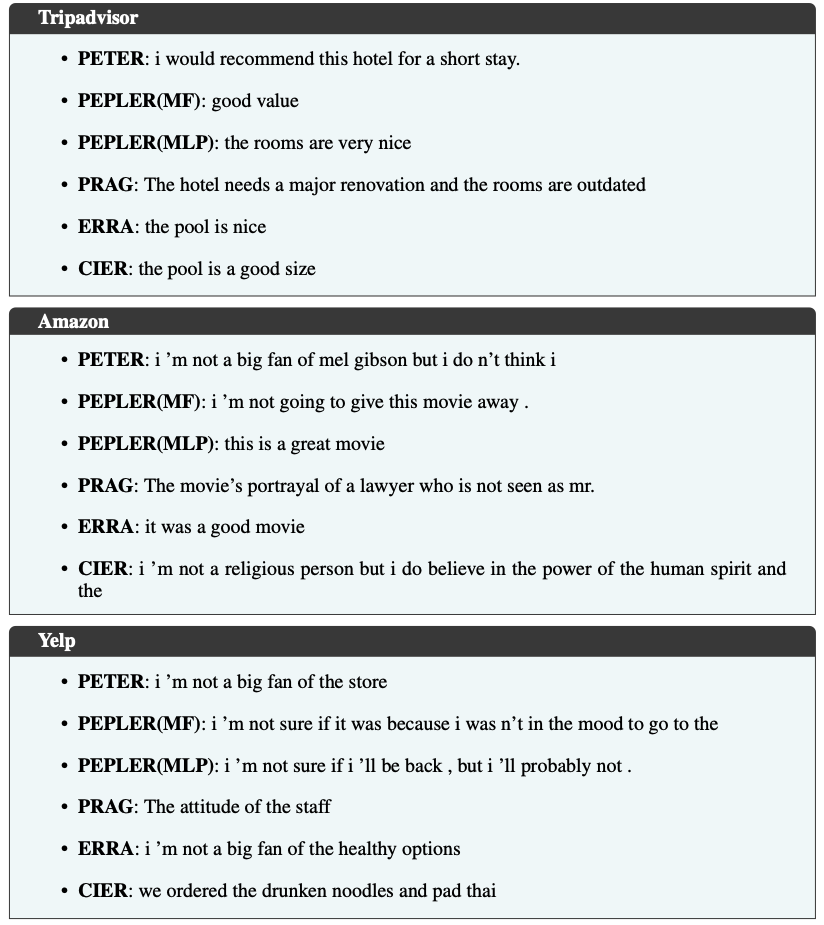}
    \caption{Examples of generated explanations from various models across Tripadvisor, Amazon, and Yelp datasets. These outputs illustrate how review-as-explanation proxies often fail to reflect the model’s actual decision-making process—frequently offering vague sentiments or unrelated opinions rather than grounded, faithful justifications.}
    \label{fig:review-not-related-decisions}
    \Description{This figure shows three examples of explanations generated by recent models for recommendation tasks on Tripadvisor, Amazon, and Yelp. Although the text in each example is fluent and grammatically correct, it often fails to mention specific user needs or item features relevant to the recommendation. Instead, the explanations are vague, overly positive, or generic—highlighting the mismatch between review-like generation and faithful justification.}
\end{figure}

Figure~\ref{fig:review-not-related-decisions} presents outputs from several recent models that adopt the review-as-explanation approach. Despite their fluency, these examples illustrate how generated explanations often fail to reflect the reasoning behind the model’s decisions—offering vague sentiments, generic praise, or irrelevant details instead of faithful justifications.

\begin{table}
    \centering
    \caption{Evaluation metrics used in prior studies on explainable recommendation systems. Each study adopts its own arbitrary set of metrics due to the absence of a systematic and unified evaluation framework.}
    \label{tab:eval_metrics}
    \begin{tabular}{@{}lccccccccccccccccc@{}}
        \toprule
        \textbf{Model} & 
        \textbf{\rotatebox{90}{BLEU}} &
        \textbf{\rotatebox{90}{ROUGE}} &
        \textbf{\rotatebox{90}{HSC}} & 
        \textbf{\rotatebox{90}{ASC}} & 
        \textbf{\rotatebox{90}{MAE}} &
        \textbf{\rotatebox{90}{RMSE}} &
        \textbf{\rotatebox{90}{USR}} &
        \textbf{\rotatebox{90}{FMR}} &
        \textbf{\rotatebox{90}{FCR}} &
        \textbf{\rotatebox{90}{DIV}} &
        \textbf{\rotatebox{90}{DISTINCT}} &
        \textbf{\rotatebox{90}{METEOR}} &
        \textbf{\rotatebox{90}{BERTScore}} &
        \textbf{\rotatebox{90}{MAUVE}} &
        \textbf{\rotatebox{90}{Entail}} & 
        \textbf{\rotatebox{90}{ENTR}} & 
        \textbf{\rotatebox{90}{ACP}} \\ 
        \midrule
        CSLARec \citeauthor{ai2025explainable}     & \checkmark & \checkmark &            &            & \checkmark & \checkmark & \checkmark & \checkmark & \checkmark & \checkmark &            &            &            &            &            &            &            \\
         SEQUER \citeauthor{ariza2023towards} & \checkmark & \checkmark &            &            & \checkmark & \checkmark & \checkmark & \checkmark & \checkmark & \checkmark &            &            &            &            &            &            &            \\
        SCRec \citeauthor{dong2025comprehend} & \checkmark & \checkmark &            &            & \checkmark & \checkmark & \checkmark & \checkmark & \checkmark & \checkmark &            &            &            &            &            &            &            \\
        NETE \citeauthor{li2020generate} & \checkmark & \checkmark &            &            & \checkmark & \checkmark & \checkmark & \checkmark & \checkmark & \checkmark &            &            &            &            &            &            &            \\
        PETER \citeauthor{li2021personalized} & \checkmark & \checkmark &            &            & \checkmark & \checkmark & \checkmark & \checkmark & \checkmark & \checkmark &            &            &            &            &            &            &            \\
        PEPLER \citeauthor{li2023personalized} & \checkmark & \checkmark &            &            & \checkmark & \checkmark & \checkmark & \checkmark & \checkmark & \checkmark &            &            &            &            &            &            &            \\
        ERRA \citeauthor{cheng2023explainable}  & \checkmark & \checkmark &            &            & \checkmark & \checkmark &            &            &            &            &            &            & \checkmark &            &            &            &            \\ 
        Att2Seq \citeauthor{dong2017learning}      & \checkmark &            & \checkmark &            &            &            &            &            &            &            &            &            &            &            &            &            &            \\
        P5 \citeauthor{geng2022recommendation} & \checkmark & \checkmark &            &            & \checkmark & \checkmark &            &            &            &            &            &            &            &            &            &            &            \\
        VIP5 \citeauthor{geng2023vip5}           & \checkmark & \checkmark &            &            &            &            &            &            &            &            &            &            &            &            &            &            &            \\
        Rexplug \citeauthor{hada2021rexplug}       & \checkmark &            &            & \checkmark &            & \checkmark &            &            &            &            & \checkmark &            &            &            &            &            &            \\
        IReGNN \citeauthor{hao2025iregnn}         &            & \checkmark &            &            &            & \checkmark &            &            &            &            &            &            &            &            &            &            &            \\
        NRT \citeauthor{li2017neural}          &            & \checkmark &            &            & \checkmark & \checkmark &            &            &            &            &            &            &            &            &            &            &            \\
        IWRER \citeauthor{li2025incorporating}   & \checkmark & \checkmark &            &            &            & \checkmark &            &            &            &            &            &            &            &            &            &            &            \\
        SERMON \citeauthor{liao2025aspect}        & \checkmark & \checkmark & \checkmark &            & \checkmark & \checkmark &  \checkmark  &            &            &            &            &            & \checkmark &            &            &            &            \\
        CIER \citeauthor{liu2025coherency} & \checkmark & \checkmark & \checkmark & \checkmark & \checkmark & \checkmark & \checkmark & \checkmark & \checkmark & \checkmark &            &            &            &            &            &            &            \\
        CER \citeauthor{Raczyński2023choherence}     & \checkmark & \checkmark & \checkmark & \checkmark & \checkmark & \checkmark & \checkmark & \checkmark & \checkmark & \checkmark &            &            &            &            &            &            &            \\
        GREENer \citeauthor{wang2022graph}         & \checkmark & \checkmark &            &            &            &            &            &            &            &            &            &            &            &            &            &            & \checkmark \\
        PRAG \citeauthor{xie2023factual}        &            &            &            &            &            & \checkmark & \checkmark &            &            &            & \checkmark &            & \checkmark & \checkmark & \checkmark & \checkmark &            \\
        D4C \citeauthor{yu2025domain}          & \checkmark & \checkmark &            &            &            &            &            &            &            &            & \checkmark & \checkmark &            &            &            &            &            \\
        \bottomrule
    \end{tabular}
\end{table}

\begin{figure}
    \centering
    \includegraphics[width=0.8\textwidth]{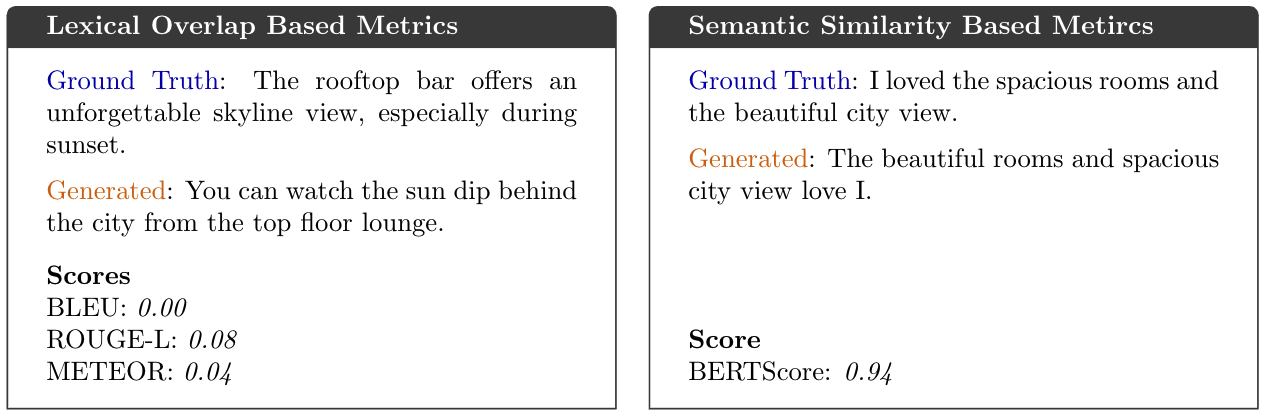}
    \caption{Limitations of standard metrics in evaluating recommendation explanations. (Left) Lexical overlap metrics such as BLEU, ROUGE, and METEOR penalize valid paraphrases despite preserving meaning. (Right) Semantic similarity metrics like BERTScore capture token-level semantic similarity but overlook syntactic structure and reasoning coherence.}
    \Description{This figure presents two illustrative cases showing why commonly used evaluation metrics may be insufficient for assessing explanation quality. On the left, two explanations with the same meaning but different wording are compared. Despite being semantically equivalent, metrics like BLEU, ROUGE, and METERO assign low scores due to limited word overlap. On the right, two explanations have similar token-level semantics, but one has wrong grammatical structure. While BERTScore gives a high similarity score, the second explanation is not valid.}
    \label{fig:metrics-problem}
\end{figure}

The review-as-explanation perspective has also shaped evaluation practices. As summarized in Table~\ref{tab:eval_metrics}, many prior studies have relied on textual overlap metrics such as BLEU and ROUGE. While these metrics are useful for measuring surface-level similarity, they are less suited for assessing semantic faithfulness. For example, they may penalize valid paraphrases while rewarding lexical overlap—even when the explanation does not reflect the model’s actual reasoning (see Figure~\ref{fig:metrics-problem}, left). Moreover, in the absence of a shared understanding of what defines a “good” explanation, evaluation practices remain diverse, with different studies adopting metrics aligned with their own assumptions.

To guide the field back toward more faithful and useful explanations, we adopt criteria inspired by recent Retrieval-Augmented Generation (RAG) literature~\cite{es2024ragas, ru2024ragchecker}, which emphasize faithfulness and relevance. We propose that effective explanations should satisfy the following principles:

\begin{enumerate}
    \item \textbf{Faithfulness} – grounded in the model’s actual decision logic.
    \item \textbf{Consistency} – sentimentally aligned with the predicted rating.
    \item \textbf{Justification} - clearly linking item features to user preferences.
    \item \textbf{Diversity} – avoiding repetition or templated expressions.
\end{enumerate}

However, many state-of-the-art methods—particularly those encoding users and items into dense embeddings as LLM tokens~\cite{li2021personalized, li2023personalized, cheng2023explainable, liu2025coherency, ariza2023towards}—struggle to meet these standards. Recommendation datasets are inherently sparse, and user preferences in reviews are often implicit, complicating semantic alignment. Consequently, generated outputs frequently miss user intent, misalign with predicted ratings, lack clear justification, or exhibit repetitive phrasing (as discussed in Section~\ref{sec:experiments}).

To address these challenges, this paper makes the following contributions:
\begin{enumerate}
    \item Review of natural language-based explainable recommendation methods, highlighting the limitations of the review-as-proxy paradigm and its influence on both modeling and evaluation practices (Section~\ref{sec:related_work}).
    \item A principled framework for evaluating explanations, grounded in faithfulness and relevance criteria from the RAG literature (Section~\ref{sec:definition}).
    \item The introduction of FIRE (\textbf{F}aithful and \textbf{I}nterpretable \textbf{R}ecommendation \textbf{E}xplanations), a framework that generates explanations by conditioning an LLM on structured inputs derived from the recommender’s most influential features. These features are identified using SHAP~\cite{shap2017, covert2021explaining, kokalj2021bert, mitchell2022gputreeshap, lundberg2020local2global}, a widely used method for model interpretability. By grounding generation in SHAP-based attributions, FIRE ensures its outputs are aligned with the actual factors behind the model’s decisions—offering a meaningful departure from review mimicry (Section~\ref{sec:proposed}).
    \item A comprehensive empirical evaluation across multiple datasets, showing how FIRE consistently produces better-aligned, more informative, and more faithful explanations compared to recent baselines (Section~\ref{sec:experiments}).
\end{enumerate}

\section{Related Studies} \label{sec:related_work}

Early efforts in generating textual explanations for recommender systems predominantly relied on template-based approaches, where fixed sentence structures were filled with explicit item attributes or topics~\cite{mcauley2013hidden, zhang2014explicit}. These methods were inherently transparent and maintained a strong connection to the underlying model logic, but lacked the flexibility to express nuanced, implicit user-item dynamics.

With the rise of deep learning, the field of explainable recommendation shifted toward free-text generation, often using user reviews as proxies for explanations. This shift began with RNN-based models utilizing LSTM architectures~\cite{dong2017learning}, and was later improved by incorporating GRUs and multi-task learning objectives—simultaneously predicting ratings and generating explanations—to enhance the robustness of learned representations~\cite{li2017neural, li2020generate}. Eventually, RNNs were replaced by Transformer-based architectures, such as PETER~\cite{li2021personalized}, where Transformer blocks were trained from scratch. More recently, models like PEPLER~\cite{li2023personalized} have integrated large pre-trained language models (PLMs), further improving generation fluency.

As the review-generation paradigm evolved, researchers began to identify several limitations—most notably, factual inconsistencies, overly generic outputs, and mismatches between the sentiment of the generated text and the model’s predicted ratings. To address factual grounding, retrieval-augmented methods such as PRAG~\cite{xie2023factual}, ERRA~\cite{cheng2023explainable}, and others~\cite{wang2022graph} incorporate relevant content from historical reviews. Concurrently, models like CIER~\cite{liu2025coherency}, REXPlug~\cite{hada2021rexplug}, and work by Raczyński et al.\cite{Raczyński2023choherence} focus on improving coherence and aligning generated sentiment with predicted scores. Building on these foundations, further refinements have explored various modeling enhancements: causal reasoning with graphs\cite{yu2025domain}, prompt-based frameworks such as P5 and VIP5~\cite{geng2022recommendation, geng2023vip5}, and integration of signals from GNNs~\cite{hao2025iregnn}, missed-review interactions~\cite{li2025incorporating}, multimodal learning~\cite{liao2025aspect}, and next-item prediction~\cite{ariza2023towards}. Additional strategies include combining summarization with efficient attention mechanisms~\cite{ai2025explainable} and two-stage prompt tuning~\cite{dong2025comprehend}.

While each of these advances targets a specific limitation—such as hallucinations, factual grounding, or sentiment alignment—most still operate under the core assumption that user reviews can function as faithful explanations. However, as demonstrated in our experimental analysis (Section~\ref{sec:experiments}), this assumption does not always hold. In many cases, explanations generated from user reviews—even when fluent or factual—remain only loosely coupled with the actual decision logic of the recommendation model.

This gap is also reflected in evaluation practices. Early models were assessed using rating prediction metrics like MSE and MAE, but with the transition to text generation, n-gram-based metrics such as BLEU, ROUGE, and METEOR~\cite{papineni2002bleu, lin-2004-rouge, banerjee-lavie-2005-meteor} became standard. These favor surface-level similarity rather than explanatory depth. Although embedding-based metrics (e.g., BERTScore~\cite{bert-score}) and distributional measures (e.g., MAUVE~\cite{pillutla-etal:mauve:neurips2021}) offer semantic improvements, they still fall short in evaluating explanation quality. This has led to the emergence of task-specific metrics targeting factuality (FMR, FCR, Entailment)\cite{li2020generate, xie2023factual}, diversity (USR, DISTINCT, n-gram Entropy)\cite{li2020generate, li2016diversity, jhamtani-etal-2018-learning}, and sentiment consistency (HSC, ASC)\cite{dong2017learning, Raczyński2023choherence, liu2025coherency}. 

Despite these developments, the field still lacks a widely accepted definition of what constitutes a “good” explanation, which contributes to fragmented evaluation practices (see Table~\ref{tab:eval_metrics}).

More recently, studies have explored using LLMs as evaluators. For example, \citet{zhang2024large} showed that LLMs can assess high-level qualities such as persuasiveness, transparency, accuracy, and user satisfaction. However, their framework stops short of clearly defining how these qualities map onto the recommendation context. In contrast, our work operationalizes these dimensions specifically for recommendation explanations. Separately, ALERT~\cite{li2025alert} proposed a source-agnostic LLM-based evaluation framework. While sharing a similar goal, our approach differs by: (1) identifying deeper issues missed by existing metrics, (2) grounding explanations in model-predicted user-item interactions, and (3) introducing a full explainable recommendation system with faithful, model-aligned outputs.

\section{Desiderata for Faithful Recommendation Explanations} \label{sec:definition}

The dominant practice in explainable recommendation has been to treat user reviews as ground-truth explanations. This framing has led to widespread reliance on metrics such as BLEU and ROUGE—tools that measure lexical overlap but are ill-equipped to assess explanatory quality. Consequently, evaluation practices across the field remain inconsistent and often misaligned with the core goal of explainability (see Table~\ref{tab:eval_metrics}). A more principled approach requires rethinking what makes a recommendation explanation effective.

To address this, we draw inspiration from recent work in the evaluation of Retrieval-Augmented Generation (RAG) systems~\cite{es2024ragas, ru2024ragchecker}. We conceptualize explanation generation in recommender systems as a RAG-like task: the system must “retrieve” (or extract) the most relevant user preferences and item attributes, and then generate a natural language explanation that faithfully answers the question, “Why is this item recommended with this predicted rating?” The RAG evaluation literature offers well-established criteria—context relevance, answer relevance, and faithfulness to the source—which serve as a rigorous foundation for defining desirable explanation properties.

Under this lens, a good explanation is not one that mimics human-written reviews, but rather one that explicitly and transparently links the model’s internal evidence to its output. For instance, a faithful positive explanation might read:\textit{``You’ll enjoy this hotel because it has a well-equipped gym, and exercise is important to you.''}
While a faithful negative explanation might state:
\textit{``You may not like this hotel as it is far from shopping centers, while proximity to malls is important to you.''}

Building on these RAG-inspired principles, we define five essential properties that a high-quality explanation $\epsilon_{u,i}$ should satisfy for a user–item pair $(u, i)$ with predicted rating $\hat{r}_{u,i}$:

\begin{enumerate}
    \item \textbf{User Alignment (Context Relevance):} The explanation should explicitly mention at least one user preference or need. This ensures relevance and personalization.
    \item \textbf{Item Alignment (Context Relevance):} It should include a concrete item feature that justifies the recommendation.
    \item \textbf{Need-Feature Connection (Answer Relevance):}  A coherent explanation must clearly link the user need to the item feature, showing how one fulfills or fails the other.
    \item \textbf{Tone-Rating Consistency (Faithfulness):} The sentiment expressed in the explanation should match the model’s predicted rating (e.g., positive tone for high ratings).
    \item \textbf{Diversity:} Explanations should be varied across instances, avoiding repetitive templates and enhancing perceived authenticity and usefulness.
\end{enumerate}
 
\section{FIRE Framework} \label{sec:proposed}

\begin{figure}
    \centering
    \includegraphics[width=0.8\textwidth]{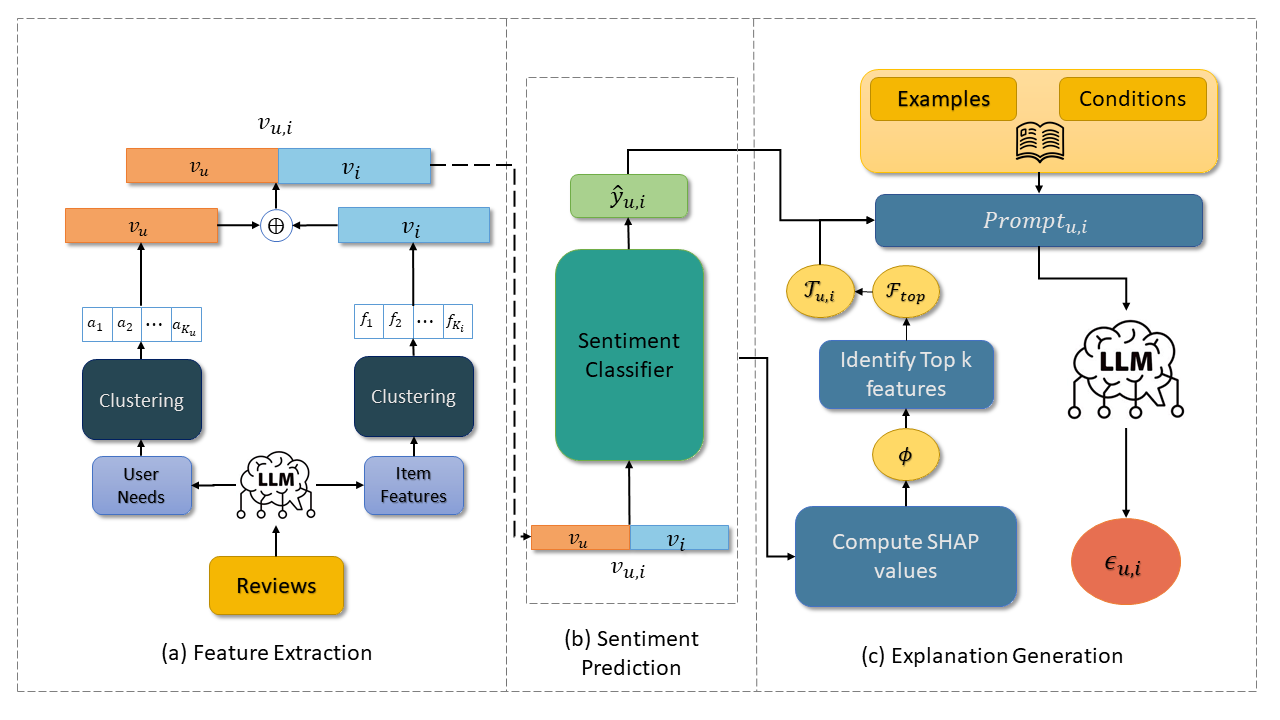}
    \caption{Overview of the FIRE framework. The pipeline includes three main stages: (a) extraction of interpretable user and item features using a pretrained language model, (b) sentiment classification using a recommendation model, and (c) explanation generation based on SHAP feature attributions and a pretrained language model.}
    \Description{illustrates the architecture of the proposed FIRE framework, which consists of three main components: (a) interpretable feature extraction using a large language model to infer user needs and item characteristics from reviews, (b) sentiment prediction via a structured recommender model trained on these interpretable features, and (c) explanation generation using SHAP-based feature attributions combined with a language model to produce faithful and human-readable justifications for the model’s predictions. The pipeline is designed to ensure that explanations are both aligned with the model’s behavior and understandable to end users.}
    \label{fig:fire}
\end{figure}

We propose FIRE, a framework designed to produce explanations that are faithful to the underlying recommendation model’s reasoning process. The pipeline comprises three stages: (1) extraction of interpretable user and item features, even when implicitly stated; (2) prediction of user preference for an item using a sentiment classification model; and (3) explanation generation based on the most influential features contributing to the model’s decision, computed via SHAP values.

\subsection{Feature Extraction}

User reviews, though inherently subjective, provide rich signals about user preferences and item characteristics. While reviews may not directly serve as faithful explanations, they offer valuable information for modeling user needs and item features.

To extract these signals, we leverage a pretrained large language model (LLM), which enables us to capture both explicit descriptions and implicit intents beyond surface-level semantics. For each ground-truth review  $e_{u,i}$ in the training set, we use an LLM\footnote{We use ‘gpt-4.1-mini’ (accessed April 2025) due to its favorable cost-performance ratio and desirable human evaluation results (see Section~\ref{sec:human-assesment}).  Other LLMs can be substituted with prompts; see Appendix~\ref{app:prompts}.}  to extract:

\begin{itemize}
    \item \textbf{User Needs:} $A(e_{u,i}) \in \mathcal{A}$, representing inferred user preferences or desires.
    \item \textbf{Item Features:} $F(e_{u,i}) \in \mathcal{F}$ , representing characteristics attributed to the item.
\end{itemize}

To manage variation in phrasing and semantic similarity, we cluster the extracted expressions into interpretable categories using unsupervised clustering. This results in two cluster sets:  $\mathcal{C}_u = \{a_1, \dots, a_{K_u}\}$ for user needs and $\mathcal{C}_i = \{f_1, \dots, f_{K_i}\}$ for item features.

We then construct structured feature vectors for users and items. Each cluster contributes two dimensions: (1) the frequency of its occurrence in a user or item’s review history, and (2) the cumulative sum of ratings associated with those mentions. The resulting vectors are:

\begin{itemize}
    \item $v_u \in \mathbb{R}^{2K_u}$: user features
    \item $v_i \in \mathbb{R}^{2K_i}$: item features
\end{itemize}

These are concatenated to form a single input vector $v_{u,i} = v_u \oplus v_i \in \mathbb{R}^{2(K_u + K_i)}$. which is used for both sentiment classification and explanation generation (see Figure~\ref{fig:fire}a).

\subsection{Sentiment Prediction}

To model user-item interactions, we adopt XGBoost~\cite{xgboost2016} due to its strong performance on structured feature data and efficient SHAP value computation. We cast the recommendation task as a three-class classification problem. Original 1–5 Likert-scale ratings $r_{u,i}$ are converted into sentiment categories as follows:

\begin{itemize}
    \item \textbf{Positive}: $r_{u, i} \geq 3.5$
    \item \textbf{Neutral}: $2.5 < r_{u, i} < 3.5$
    \item \textbf{Negative}: $r_{u, i} \leq 2.5$
\end{itemize}

This reformulation is motivated by the observation that natural language explanations do not require exact numeric ratings, but rather a reliable indication of the user’s attitude toward the item. The model is trained using standard cross-entropy loss:

\begin{equation}
    \mathcal{L} = - \sum_{j = 1}^N  \sum_{c = 1}^3 y_j^c \log \hat{y}_j^c
\end{equation}

where $y_j^c$ is the ground-truth one-hot label and $\hat{y}_j^c$ the predicted class probability for the $j^{\text{th}}$ interaction (Figure~\ref{fig:fire}b).

\subsection{Explanation Generation }

\begin{algorithm}[tb]
\caption{Explanation Generation}
\label{alg:explanation}
\textbf{Input}: Trained model $M$, feature vector $v_{u,i}$, SHAP explainer $\mathcal{S}$, explanation generator $\mathcal{G}$
\begin{algorithmic}[1]
\STATE $\hat{y}_{u,i} \gets M(v_{u,i})$ \COMMENT{Model prediction}
\STATE $\phi \gets \mathcal{S}(M, v_{u,i})$ \COMMENT{Compute SHAP values}
\STATE $\mathcal{F}_{top} \gets$ Top-$K$ features with highest $|\phi|$
\STATE $\mathcal{T}_{u,i} \gets$ Map $\mathcal{F}_{top}$ to cluster descriptions
\STATE $\mathcal{P}_{u,i}(\mathcal{T}_{u,i}, \hat{y}_{u,i}, \text{conditions}, \text{few-shot examples})$
\STATE $\epsilon_{u,i} \gets \mathcal{G}(\mathcal{P}_{u,i})$ \\
\STATE \textbf{return} $\epsilon_{u,i}$
\end{algorithmic}
\end{algorithm}

To ensure that explanations reflect the recommendation model’s actual decision logic, we use SHAP (SHapley Additive exPlanations)\cite{shap2017}. SHAP is a game-theoretic method that assigns an importance value to each feature based on its contribution to the model’s prediction. It is widely adopted across machine learning domains for feature attribution\cite{covert2021explaining, kokalj2021bert, mitchell2022gputreeshap, lundberg2020local2global}.

To generate explanations that are faithful to the model’s decision-making process, given a trained model $M$ and a user–item feature vector $v_{u,i}$, we first compute the predicted sentiment $\hat{y}_{u,i} = M(v_{u,i})$. Next, we use a SHAP explainer $\mathcal{S}$ to obtain a set of importance scores $\phi = \mathcal{S}(M, v_{u,i})$, where each score reflects the contribution of an input feature to the prediction. We then identify the top-$K$ features with the highest absolute SHAP values, representing the most influential factors in the model’s decision. These features are mapped back to their corresponding human-readable cluster descriptions, denoted $\mathcal{T}_{u,i}$. Using this set of influential features, along with the predicted sentiment $\hat{y}_{u,i}$ and a few-shot template, we construct a prompt $\mathcal{P}_{u,i}$ designed to elicit a coherent explanation. Finally, this prompt is passed to a pretrained large language model $\mathcal{G}$, which generates the textual explanation $\epsilon_{u,i}$. This explanation reflects the internal reasoning of the model while remaining interpretable and informative to end users.

This process is outlined in Algorithm~\ref{alg:explanation} and illustrated in Figure~\ref{fig:fire}c.

\section{Experiments}
\label{sec:experiments}

We evaluate FIRE against representative state-of-the-art (SOTA) models across the dimensions of desirable explanation characteristics introduced in Section~\ref{sec:definition}. Our experiments aim to answer the following research questions:

\begin{enumerate}
    \item \textbf{RQ1:} How does FIRE’s recommendation performance compare to contemporary SOTA methods?
    \item \textbf{RQ2:}  Does FIRE’s design—eschewing the “review-as-proxy” paradigm—yield explanations that better reflect the targeted desirable traits?
    \item \textbf{RQ3:} To what extent do FIRE’s LLM-based explanations and evaluation outputs align with human judgments?
\end{enumerate}

\subsection{Dataset}
\label{sec:dataset}

\begin{table}[t]
\centering
\caption{Dataset Statistics}
\label{tab:dataset_stats}
\begin{tabular}{lccc}
\toprule
\textbf{Metric} & \textbf{Amazon} & \textbf{Yelp} & \textbf{TripAdvisor} \\
\midrule
\#Users & 7506 & 27147 & 9765 \\
\#Products & 7360 & 20266 & 6280 \\
\#Records & 441783 & 1293247 & 320203 \\
Records/User & 58.86 & 47.64 & 32.77 \\
Records/Product & 60.02 & 63.81 & 50.96 \\
Average Review Length & 14.05 & 12.23 & 12.98 \\
Class Ratio (Pos:Neu:Neg) & 4.8:1.3:1 & 5.4:1.1:1 & 14.7:3.1:1 \\
\bottomrule
\end{tabular}
\end{table}

We evaluate our framework using three widely adopted benchmark datasets from distinct domains: TripAdvisor (hotels), Amazon (movies and TV), and Yelp (restaurants). These datasets were originally introduced by~\cite{li2020generate} and have since been extensively used in subsequent studies on natural language explainable recommendation~\cite{li2021personalized, li2023personalized, xie2023factual, cheng2023explainable}\footnote{\url{https://github.com/lileipisces/PETER/tree/master}}. Together, they cover diverse application contexts and exhibit varying levels of data sparsity, enabling a comprehensive assessment of explanation quality under different recommendation scenarios.

To ensure alignment with prior work, we retain users and items with at least 20 interactions. Each dataset is partitioned into training, validation, and test sets using an 8:1:1 ratio, while guaranteeing that every user and item is represented in the training set.

Table~\ref{tab:dataset_stats} summarizes key dataset statistics. The datasets differ in size, interaction density, and average review length. Notably, all three exhibit a pronounced class imbalance, with a significantly higher proportion of positive reviews relative to neutral and negative ones—posing additional challenges for generating balanced and meaningful explanations.

\subsection{Baselines}
\label{sec:baselines}

We evaluate the performance of our proposed FIRE framework by comparing it against several state-of-the-art baselines spanning three major categories of explanation generation methods:

(1) \textbf{Language model-based generators:} We include PETER~\cite{li2021personalized} and PEPLER~\cite{li2023personalized}, both of which utilize pre-trained language models to generate personalized reviews. For PEPLER, we consider both matrix factorization (MF) and multilayer perceptron (MLP) variants to reflect different collaborative modeling strategies.

(2) \textbf{Retrieval-augmented models:} We compare against ERRA~\cite{cheng2023explainable} and PRAG~\cite{xie2023factual}, which enhance generation quality and factual consistency by retrieving relevant historical reviews during the decoding process.

(3) \textbf{Sentiment-aware generators:} We include CIER~\cite{liu2025coherency}, a model that aligns the sentiment of generated explanations with the predicted rating.

\subsection{Recommendation Performance}

\begin{table}
\centering
\caption{F1-scores of different models across three datasets}
\label{tab:f1_scores}
\begin{tabular}{lccc}
\toprule
\textbf{Model} & \textbf{TripAdvisor} & \textbf{Amazon} & \textbf{Yelp} \\
\midrule
PETER          & 0.73  & 0.71  & 0.63 \\
PEPLER (MF)    & 0.31  & 0.58  & 0.47 \\
PEPLER (MLP)   & 0.74  & 0.71  & 0.62 \\
PRAG           & 0.71  & 0.72  & 0.63 \\
ERRA           & 0.74  & 0.72  & 0.63 \\
CIER           & 0.74  & 0.70  & 0.62 \\
FIRE           & 0.73  & 0.72  & 0.65 \\
\bottomrule
\end{tabular}
\end{table}

Given the imbalanced nature of the rating distributions in all three datasets, we report weighted F1-scores to ensure fair comparison across models. The results are presented in Table~\ref{tab:f1_scores}.

FIRE achieves the best performance on the Yelp and Amazon datasets, while remaining highly competitive on TripAdvisor, closely matching or outperforming several deep learning-based baselines. 

These findings highlight an important insight: high-quality recommendation performance can be achieved without relying on complex, token-based dense embeddings tightly embedded within deep model architectures. Instead, FIRE reaches competitive accuracy by leveraging interpretable features and a lightweight XGBoost model. This design allows for fast, resource-efficient training—often completed in minutes on a single GPU—while offering transparency in the prediction process, a crucial factor for building explainable and trustworthy systems.

\subsection{Rating–Sentiment Consistency in Explanations} \label{sec:consistency}

An effective explanation should align with the sentiment of the corresponding rating—positive reviews for high scores, negative for low. Although many models jointly train on both rating prediction and text generation, they often fail to maintain this alignment due to the lack of explicit conditioning on the predicted sentiment~\cite{li2020generate, li2021personalized, li2023personalized, cheng2023explainable}.

To assess sentiment consistency, we evaluate whether the tone of a generated explanation matches the sentiment of its predicted rating. Each predicted rating (positive, neutral, or negative) is paired with its corresponding explanation, and we use a robust sentiment classifier~\cite{hartmann2023} to estimate the sentiment of the explanation text. The classifier returns a score between 0 (strongly negative) and 1 (strongly positive), reflecting the explanation’s sentiment polarity.

\begin{figure}
    \centering
    \includegraphics[width=0.8\textwidth]{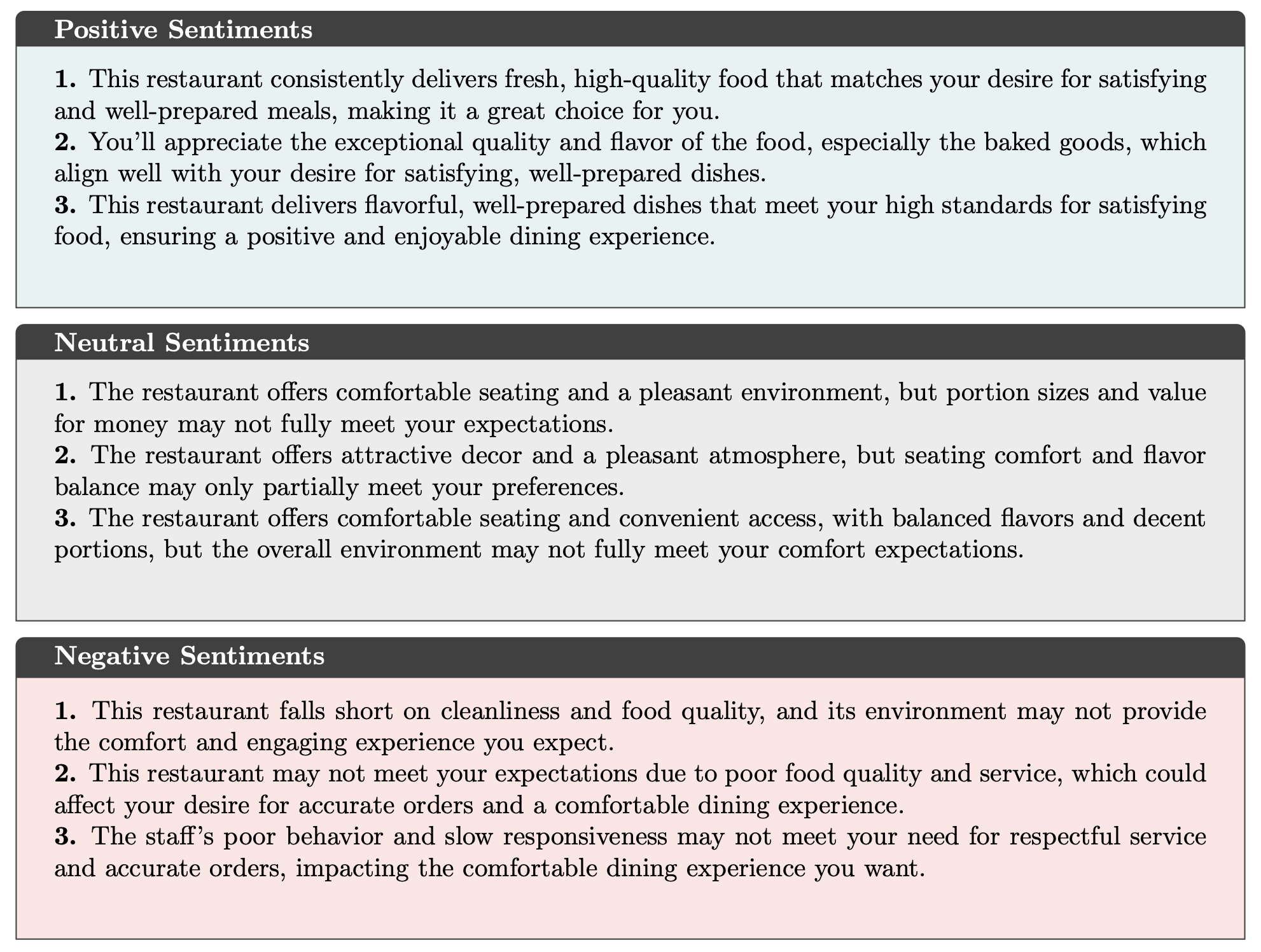}
    \caption{Examples of explanations generated by FIRE, categorized into positive, neutral, and negative sentiment classes.}
    \Description{Three blocks of text demonstrate how FIRE adapts tone based on predicted sentiment, highlighting its ability to generate emotionally aligned explanations across sentiment classes.}
    \label{fig:sentiment-consistency}
\end{figure}

Figure~\ref{fig:sentiment-consistency} presents illustrative examples of generated explanations across sentiment categories. These samples show that FIRE generates explanations with clear and appropriate emotional tone corresponding to each predicted rating, demonstrating a key aspect of faithfulness in explainable recommendation.

\begin{figure}
    \centering
    \includegraphics[width=0.8\textwidth]{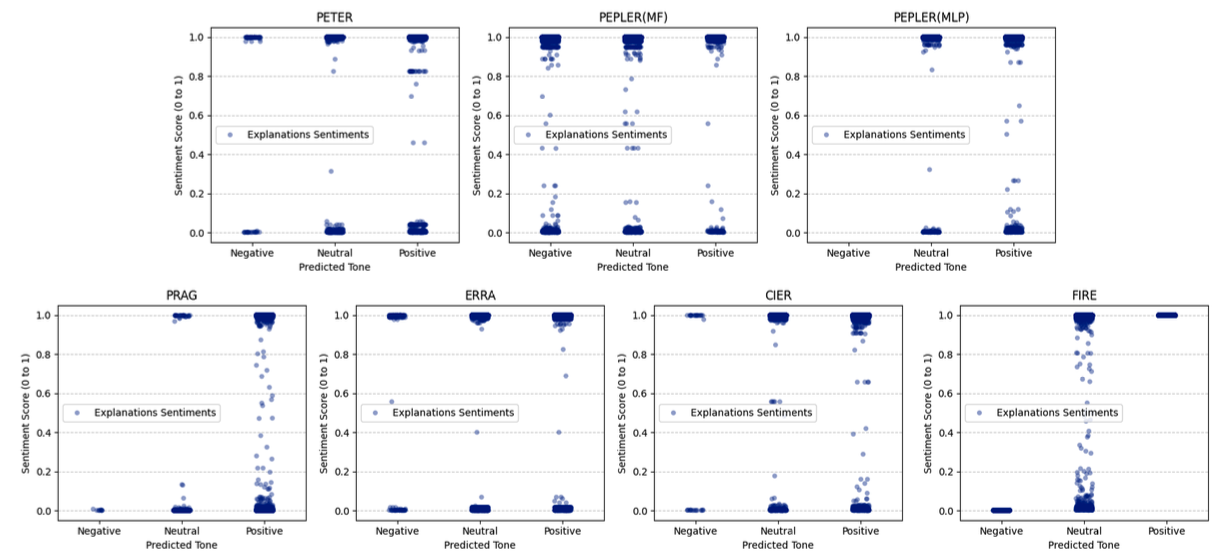}
    \caption{Sentiment consistency of generated explanations vs. predicted ratings on the TripAdvisor dataset.}
    \Description{A scatter plot showing predicted ratings on the x-axis and sentiment scores on the y-axis. A clear upward trend indicates strong alignment between rating predictions and explanation tone.}
    \label{fig:rating-concistentcy-tripadvisor}
\end{figure}

\begin{figure}
    \centering
    \includegraphics[width=0.8\textwidth]{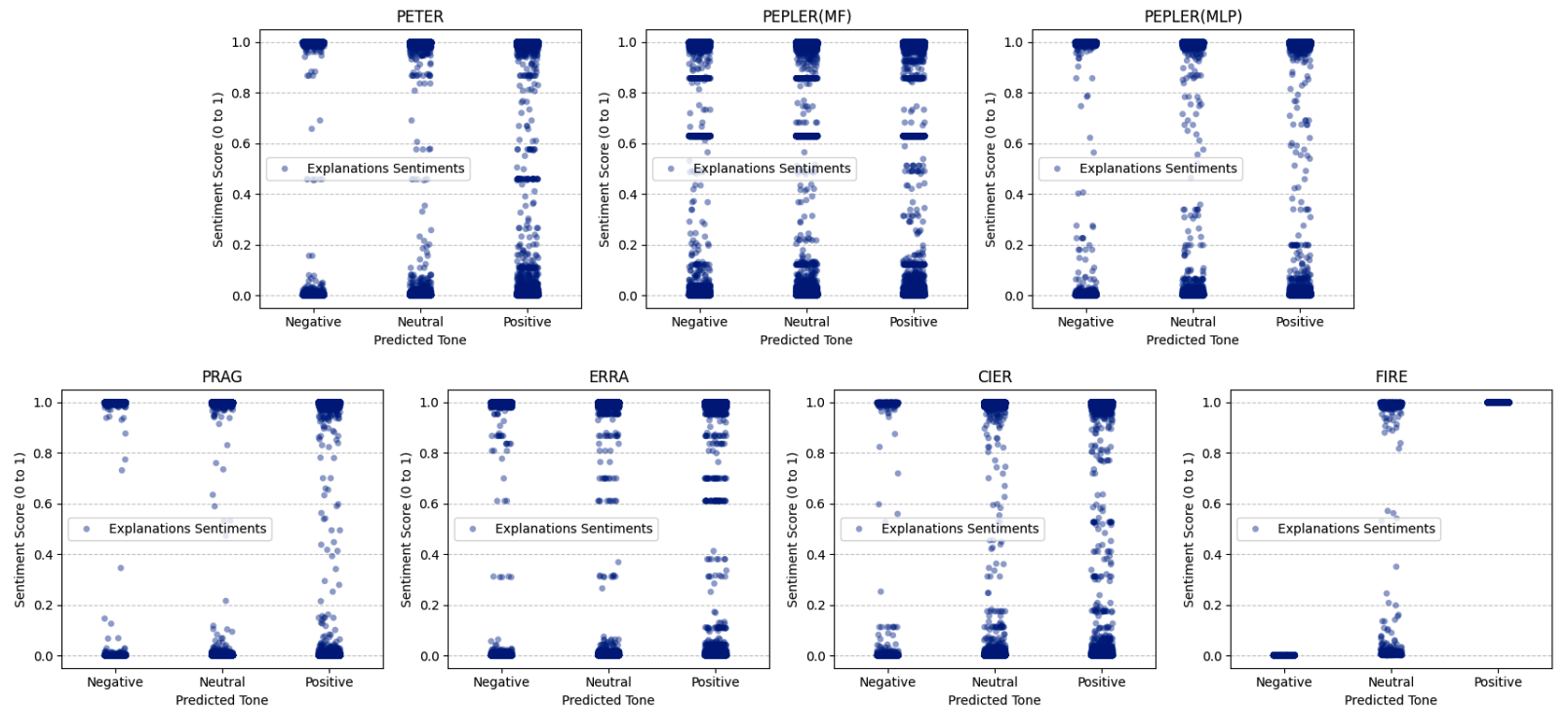}
    \caption{Sentiment consistency of generated explanations vs. predicted ratings on the Amazon dataset.}
    \Description{Similar to the TripAdvisor plot, points are concentrated in the upper-right for positive ratings and bottom-left for negative ratings, showing consistent alignment.}
    \label{fig:rating-concistentcy-amazon}
\end{figure}

\begin{figure}
    \centering
    \includegraphics[width=0.8\textwidth]{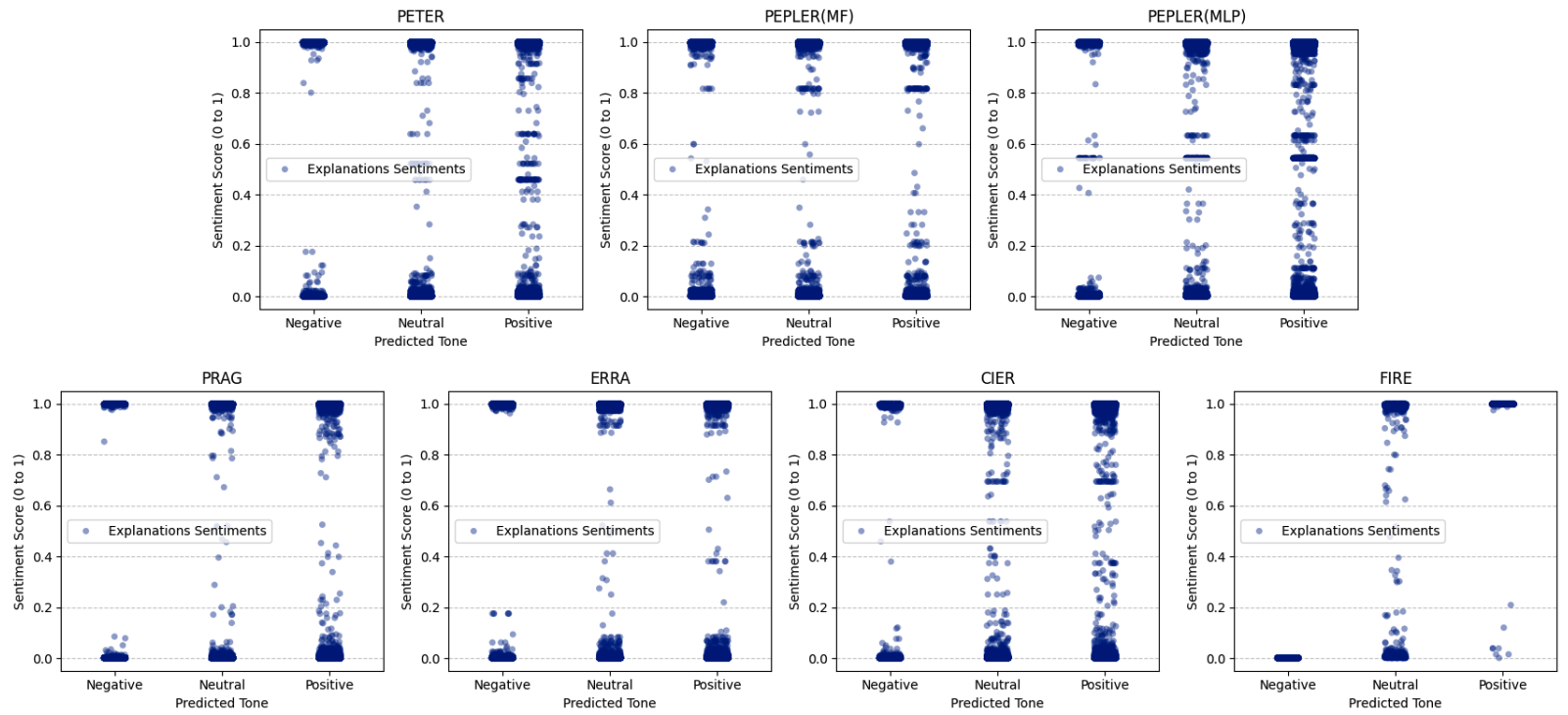}
    \caption{Sentiment consistency of generated explanations vs. predicted ratings on the Yelp dataset.}
    \Description{The scatter plot illustrates sentiment scores rising with predicted ratings. FIRE maintains a strong correlation, particularly in the presence of more neutral reviews compared to other datasets.}
    \label{fig:rating-concistentcy-yelp}
\end{figure}

Figures~\ref{fig:rating-concistentcy-tripadvisor}, \ref{fig:rating-concistentcy-amazon}, and \ref{fig:rating-concistentcy-yelp} present sentiment consistency scatter plots for the TripAdvisor, Amazon, and Yelp datasets, respectively. Each dot represents a test instance, with the x-axis showing the predicted rating and the y-axis showing the sentiment score of the generated explanation. A strong upward trend reflects good alignment.

While some baselines show partial consistency—PRAG, for instance, performs relatively well for negative ratings on TripAdviosr due to its question-answering generation style (“What was not good?”)—most models fail to enforce alignment across the full rating range. Many models treat rating and explanation generation as multi-task objectives but don’t explicitly condition the generated text on the rating output, leading to mismatches in tone.

Even CIER, a model designed to improve textual coherence, struggles with sentiment misalignment. Their evaluation metric, which considers a prediction consistent if its sentiment is within one star of the true rating, is overly lenient—especially problematic given the class imbalance in recommendation data, where positive ratings dominate. This coarse-grained measure can mask inconsistencies and inflate performance metrics.

In contrast, FIRE maintains strong alignment between predicted sentiment and explanation tone across datasets. As shown in the scatter plots, explanations for negative ratings consistently receive low sentiment scores (near 0), those for neutral ratings vary more widely, and those for positive ratings cluster near 1.

This alignment is intentional: FIRE conditions its explanation generation prompt explicitly on the predicted sentiment (see Appendix~\ref{app:prompts}). This prompt-driven design enables FIRE to produce coherent, sentiment-aligned explanations, ensuring consistency.

\subsection{User–Item Alignment in Explanations} \label{sec:alignment}

For an explanation to be faithful, it must reflect the reasoning behind the recommendation—specifically, by referencing the user’s need and the relevant item feature referenced in source text. Despite this being a core requirement for explainability, many recent models fail to meet it, often generating fluent but generic explanations that do not align with the underlying decision logic, even implicitly.

\begin{figure}
    \centering
    \includegraphics[width=0.8\textwidth]{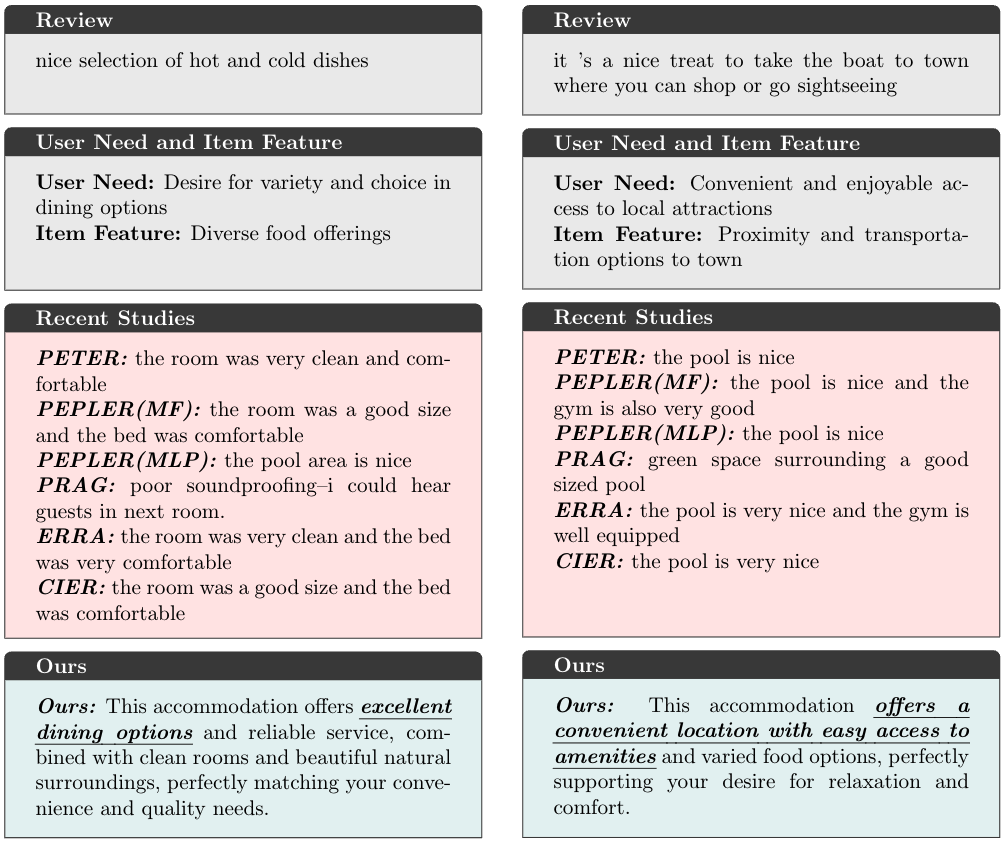}
    \caption{Examples comparing outputs of different models to ground-truth reviews, highlighting how generated explanations often fail to mention the user’s need or relevant item features, even implicitly.}
    \label{fig:alignment_examples}
    \Description{This figure shows how prior models often fail to mention relevant user needs or item features, even when such rationale is present in the true review. In contrast, FIRE accurately identifies and integrates both aspects into the explanation.}
\end{figure}

As illustrated in Figure~\ref{fig:alignment_examples}, explanations generated by several existing methods often fail to identify the correct user needs or item features that justify the recommendation. In contrast, the FIRE framework captures the appropriate user need and item feature and integrates them into the explanation. This demonstrates FIRE’s ability to align closely with the model’s prediction while grounding its reasoning in the same key elements present in the original review.

To evaluate this alignment systematically, we extract user needs ($\mathcal{A}$) and item features ($\mathcal{F}$) from ground-truth reviews $e_{u,i}$ using a pretrained LLM, and compare them to the same elements in generated explanations ($\epsilon_{u,i}$), computing alignment scores $S_u$ and $S_i$:

\begin{equation}
    \begin{split}
        & S_u = \frac{1}{|\mathcal{D}_\text{test}|} \sum_{e_{u, i} \in \mathcal{D}_{\text{test}}} \mathbb{I}[A(e_{u, i}) = A(\epsilon_{u, i})] \\
        & S_i = \frac{1}{|\mathcal{D}_\text{test}|} \sum_{e_{u, i} \in \mathcal{D}_{\text{test}}} \mathbb{I}[F(e_{u, i}) = F(\epsilon_{u, i})] \\
    \end{split}
\end{equation}

\begin{table}
\caption{Alignment scores of user needs and item features across three datasets.}
\label{tab:alignment_scores}
\centering
\begin{tabular}{ccccccccc}
\toprule
 &  & \textbf{PETER} & \textbf{PEPLER(MF)} & \textbf{PEPLER(MLP)} & \textbf{PRAG} & \textbf{ERRA} & \textbf{CIER} & \textbf{FIRE} \\ \hline
\multirow{2}{*}{\textbf{TripAdvisor}} & $S_u$ & 0.31 & 0.32 & 0.3 & \underline{0.46} & 0.31 & 0.32  & \textbf{0.56}\\
                                      & $S_i$ & 0.14 & 0.16 & 0.15 & \underline{0.22} & 0.15 & 0.16 & \textbf{0.24} \\ \hline
\multirow{2}{*}{\textbf{Amazon}}      & $S_u$ & 0.24 & 0.28 & 0.31 &\underline{0.48} & 0.25 & 0.32 & \textbf{0.57}\\
                                      & $S_i$ & 0.13 & 0.20 & 0.18 & \underline{0.36} & 0.14 & 0.18 & \textbf{0.50} \\ \hline
\multirow{2}{*}{\textbf{Yelp}}        & $S_u$ & 0.35 & 0.31 & 0.37 & \underline{0.50} & 0.35 & 0.36 & \textbf{0.50}\\
                                      & $S_i$ & 0.17 & 0.15 & 0.19 & \textbf{0.29} & 0.17 & 0.18 & \underline{0.26}\\
\bottomrule
\end{tabular}
\end{table}

Table~\ref{tab:alignment_scores} presents alignment scores across three datasets. Most baseline models exhibit poor alignment with user and item rationales, highlighting a lack of faithfulness in their generated explanations.

A key reason for low alignment is the reliance on dense token-level user/item embeddings, which lack the granularity to surface specific user needs or item features—especially when these are not overtly mentioned in training reviews. As a result, many models fail to reflect what truly drove the prediction.

One exception is PRAG, which performs better on both alignment metrics by retrieving and conditioning on past reviews using keyword extraction and a QA-style generation model. Its retrieval-based design allows it to surface more concrete item features, especially for negative or nuanced cases.

Despite PRAG’s advantages, FIRE consistently outperforms all baselines. This performance stems from two critical design decisions:
\begin{enumerate}
    \item Interpretable input features mapped from review-derived semantic clusters, making the model aware of meaningful concepts.
    \item SHAP-based explanation generation, which grounds each explanation in the actual predictive features used by the model, rather than simply mimicking review style or content.
\end{enumerate}

Together, these mechanisms ensure that FIRE’s explanations are not only fluent, but also aligned with the internal logic of the recommendation—an essential property for building trust in real-world applications.

\subsection{Structure in Textual Explanations} \label{sec:format}

\begin{figure}
    \centering
    \includegraphics[width=0.8\textwidth]{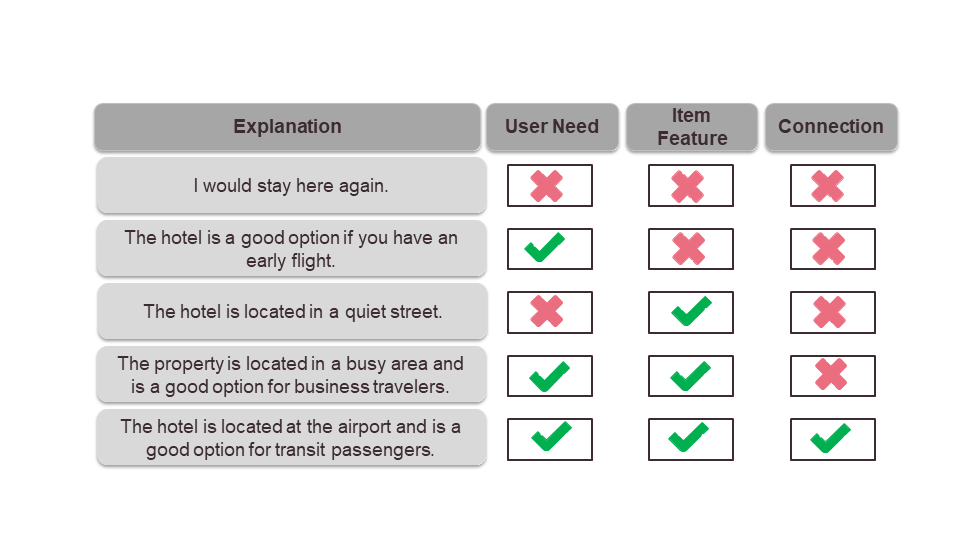}
    \caption{Example outputs from CIER on the TripAdvisor dataset showing how review-generation-based methods often fail to follow the desired explanation format.}
    \Description{Example explanations from the CIER model on the TripAdvisor dataset. The figure illustrates how omitting user needs, item features, or the connection between them leads to ineffective explanations. The final row shows a well-structured output including all essential components.}
    \label{fig:format-examples}
\end{figure}

A high-quality explanation in a recommender system should be structured and rationale-driven. Specifically, it must communicate:

\begin{itemize}
    \item What the model believes about the user (i.e., the user’s needs),
    \item What it knows about the item (i.e., the item’s features), and
    \item Why it predicts the item to be a good or poor match, by connecting those two aspects.
\end{itemize}

However, many existing explanation models fail to provide this structure. As shown in Figure~\ref{fig:format-examples}, generated outputs from the CIER model often omit either user needs, item features, or a logical bridge between them—undermining the explanation’s clarity and faithfulness. Only the final row illustrates a well-structured explanation that includes all critical elements.

To evaluate structure systematically, we used a pretrained large language model to assess each explanation across three key criteria: (1) Presence of a user need, (2) Presence of an item feature, and (3) A logical connection between the two.

\begin{figure}
    \centering
    \includegraphics[width=0.8\textwidth]{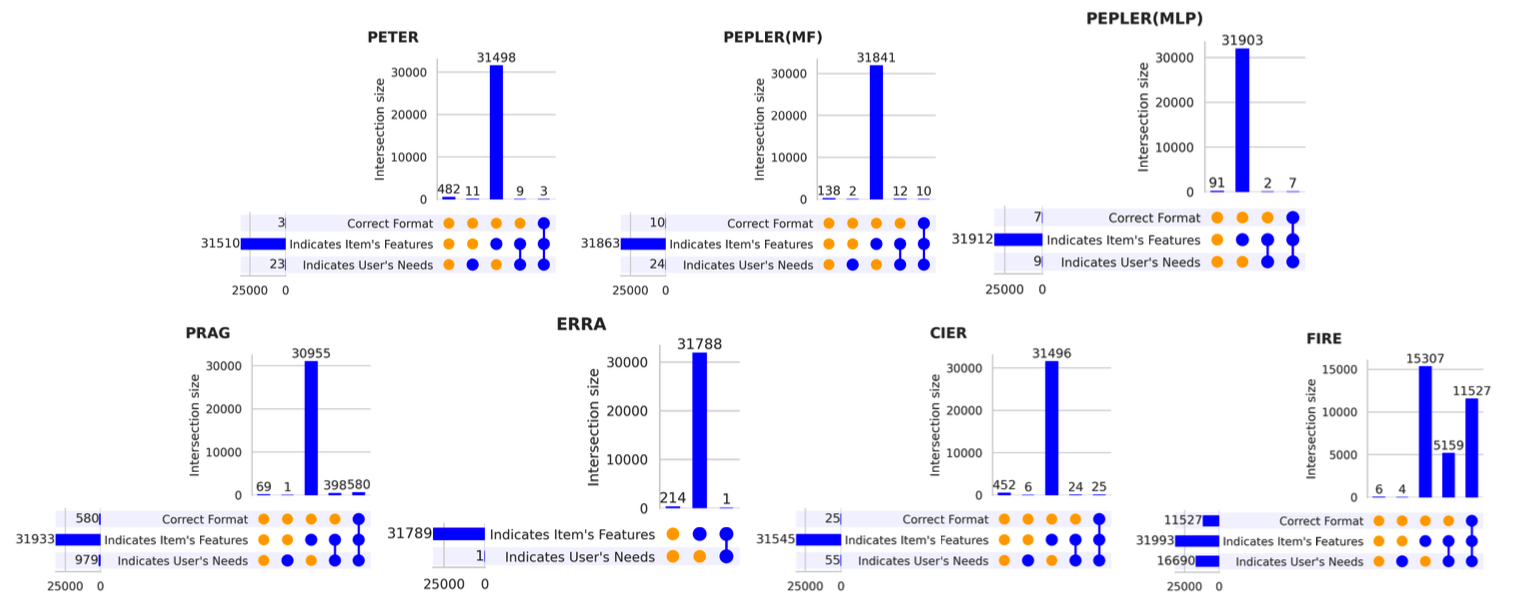}
    \caption{Results of explanation format analysis on the TripAdvisor dataset. Outputs from recent explainable recommender models largely fail to meet the structural criteria for proper explanations.}
    \Description{Evaluation of explanation structure adherence for TripAdvisor. FIRE significantly outperforms other models in generating structured, rationale-based outputs.}
    \label{fig:format-tripadvisor}
\end{figure}

\begin{figure}
    \centering
    \includegraphics[width=0.8\textwidth]{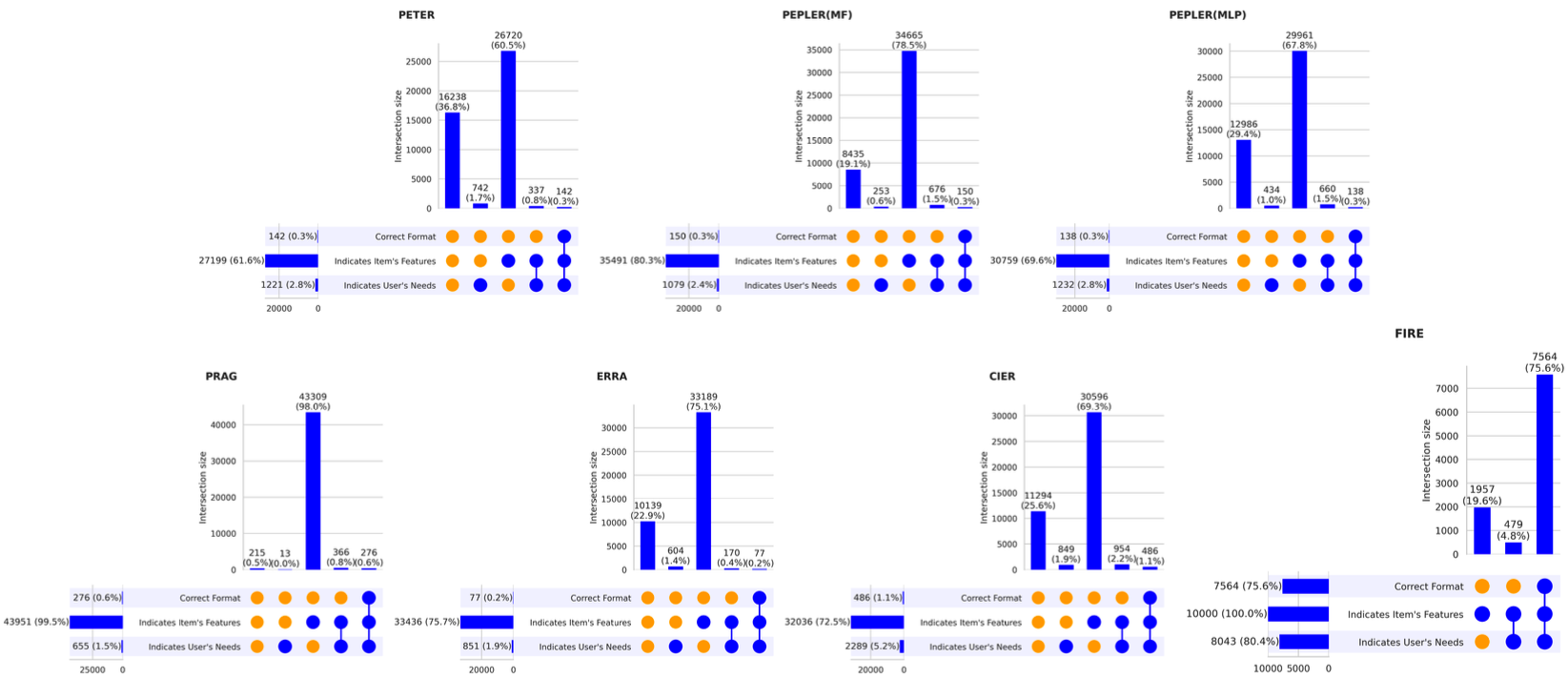}
    \caption{ Results of explanation format analysis on the Amazon dataset. Outputs from recent explainable recommender models largely fail to meet the structural criteria for proper explanations.}
    \Description{Structural evaluation results on the Amazon dataset. Similar to TripAdvisor, most models fail to include all required elements in their explanations.}
    \label{fig:format-amazon}
\end{figure}

\begin{figure}
    \centering
    \includegraphics[width=0.8\textwidth]{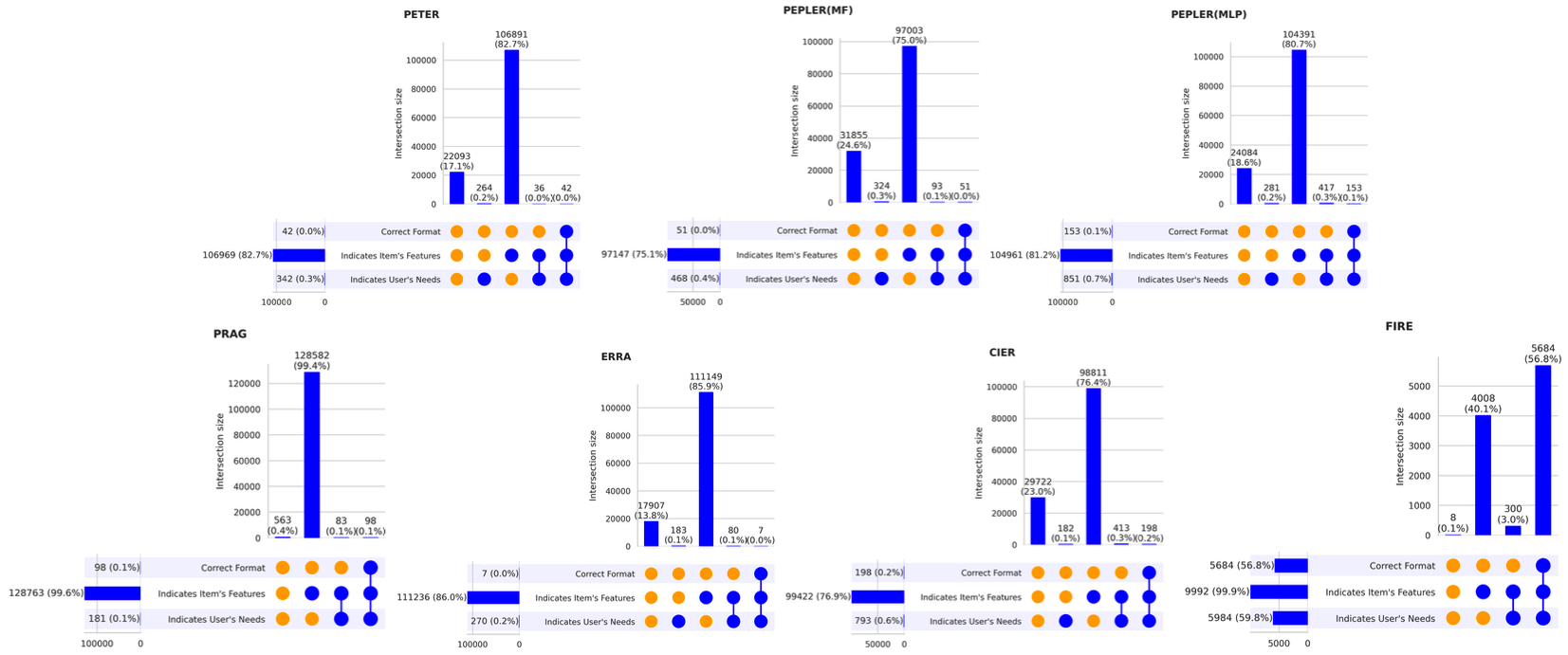}
    \caption{ Results of explanation format analysis on the Yelp dataset. Outputs from recent explainable recommender models largely fail to meet the structural criteria for proper explanations.}
    \Description{Structural analysis results on the Yelp dataset, again confirming that FIRE consistently produces well-structured justifications while baselines struggle.}
    \label{fig:format-yelp}
\end{figure}

This evaluation was applied to test-set outputs from all models across the TripAdvisor, Amazon, and Yelp datasets. The results are presented in Figures~\ref{fig:format-tripadvisor}, \ref{fig:format-amazon}, and \ref{fig:format-yelp}.

Across all datasets, most baseline methods struggled to generate explanations with clear structure. This limitation arises from the review-as-proxy paradigm, which attempts to replicate user-written reviews rather than generating targeted justifications. Reviews tend to lack explicit intent or reasoning, as users write them retrospectively rather than explain proactively.


FIRE’s superiority stems from its prompting strategy, which explicitly instructs the model to generate structured explanations that (i) highlight user needs, (ii) cite item features, and (iii) form a coherent link between the two. This structure results in explanations that are interpretable, grounded, and faithful to the model’s decision-making process.

\subsubsection{Limitation}

\begin{figure}
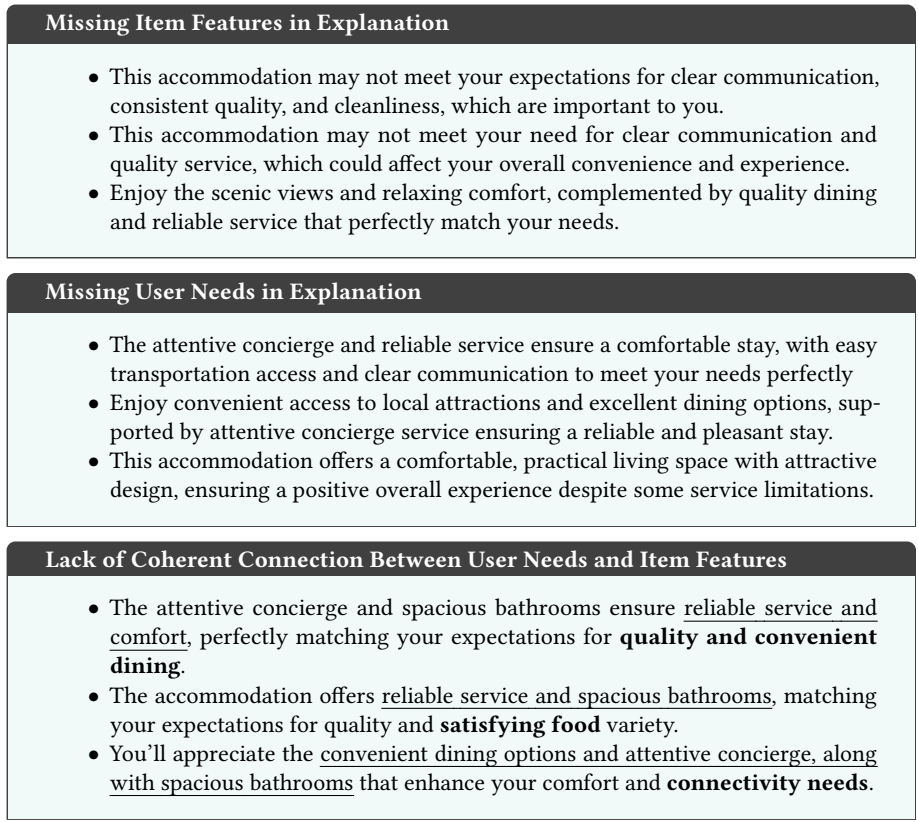

    \centering
    \begin{tcolorbox}[colback=teal!5, colframe=black!75!white, title=Missing Item Features in Explanation, fonttitle=\bfseries, width=0.8\textwidth, boxrule=0.5pt, sharp corners=south]
    \begin{itemize}
        \item This accommodation may not meet your expectations for clear communication, consistent quality, and cleanliness, which are important to you.
        \item This accommodation may not meet your need for clear communication and quality service, which could affect your overall convenience and experience.
        \item Enjoy the scenic views and relaxing comfort, complemented by quality dining and reliable service that perfectly match your needs.
    \end{itemize}
    \end{tcolorbox}
    \begin{tcolorbox}[colback=teal!5, colframe=black!75!white, title=Missing User Needs in Explanation, fonttitle=\bfseries, width=0.8\textwidth, boxrule=0.5pt, sharp corners=south]
    \begin{itemize}
        \item  The attentive concierge and reliable service ensure a comfortable stay, with easy transportation access and clear communication to meet your needs perfectly
        \item Enjoy convenient access to local attractions and excellent dining options, supported by attentive concierge service ensuring a reliable and pleasant stay.
        \item This accommodation offers a comfortable, practical living space with attractive design, ensuring a positive overall experience despite some service limitations.
    \end{itemize}
    \end{tcolorbox}
    \begin{tcolorbox}[colback=teal!5, colframe=black!75!white, title=Lack of Coherent Connection Between User Needs and Item Features, fonttitle=\bfseries, width=0.8\textwidth, boxrule=0.5pt, sharp corners=south]
    \begin{itemize}
        \item The attentive concierge and spacious bathrooms ensure \uline{reliable service and comfort}, perfectly matching your expectations for \textbf{quality and convenient dining}.
        \item The accommodation offers \uline{reliable service and spacious bathrooms}, matching your expectations for quality and \textbf{satisfying food} variety.
        \item You'll appreciate the \uline{convenient dining options and attentive concierge, along with spacious bathrooms} that enhance your comfort and \textbf{connectivity needs}.
    \end{itemize}
    \end{tcolorbox}
    \caption{Examples of failure cases in FIRE-generated explanations on the Tripadvisor dataset. Top: Explanations lacking item features. Center: Explanations lacking user needs. Bottom: Explanations where the user needs and item features are not meaningfully connected.}
    \Description{Illustrative failure cases from FIRE-generated explanations. Top box shows missing item features, center shows missing user needs, and bottom reveals examples where present elements fail to form a coherent narrative.}
    \label{fig:failure-examples}
\end{figure}

Despite its strengths, FIRE occasionally fails to produce fully structured explanations. Figure~\ref{fig:failure-examples} illustrates three common types of failure cases:
\begin{itemize}
    \item \textbf{Missing Item Features:} When all influential features belong to user clusters, the explanation lacks item-specific justifications.
    \item \textbf{Missing User Needs:} When influential features are item-based, the user’s motivations are omitted.
    \item \textbf{Disconnected Components:} Even when both user and item aspects are present, no explanation could logically connect them, reducing coherence and persuasiveness.
\end{itemize}

These limitations highlight areas for future improvement, though FIRE still sets a strong benchmark for explanation structure in recommendation.

\subsection{Sentence Variation in Model Outputs} \label{sec:diversity}

Linguistic diversity is a hallmark of human-written reviews. To capture this quality in generated explanations, we evaluate models using the Unique Sentence Ratio (USR)—the proportion of unique sentences among all generated outputs:

\begin{equation}
    \text{USR} = \frac{|\{\hat{e}^s\}_{s=1}^N|}{N}
\end{equation}

where $|\{\hat{e}^s\}_{s=1}^N|$ is the number of unique generated sentences and $N$ is the total number of generated explanations. A higher USR indicates greater variation and is desirable, as long as explanatory quality is maintained.

\begin{table}
\centering
\caption{USR scores across different models and datasets.}
\label{tab:usr}
\begin{tabular}{lccc}
\toprule
\textbf{Model} & \textbf{TripAdvisor} & \textbf{Amazon} & \textbf{Yelp} \\
\midrule
Dataset        & 0.98 & 0.98 & 0.98 \\ \hline
PETER          & 0.08 & 0.31 & 0.10  \\
PEPLER (MF)    & 0.21 & 0.32 & 0.09  \\
PEPLER (MLP)   & 0.19 & 0.39 & 0.21  \\
PRAG           & \underline{0.97} & \underline{0.91} & \underline{0.77} \\
ERRA           & 0.05 & 0.15 & 0.10  \\
CIER           & 0.30 & 0.53 & 0.30 \\
FIRE           & \textbf{0.99} & \textbf{0.91} & \textbf{0.95}\\
\bottomrule
\end{tabular}
\end{table}

Table~\ref{tab:usr} presents USR scores across models and datasets. Human-written reviews naturally achieve near-perfect diversity (0.98 USR). In contrast, many recent models—including PETER, PEPLER, ERRA, and CIER—produce highly repetitive outputs, with low USR scores.

This lack of variation can be traced back to the models’ reliance on dense user/item embeddings learned from sparse data. These representations  fail to capture fine-grained user preferences and item distinctions—resulting in repetitive and semantically limited outputs.

Among baselines, PRAG performs relatively well, thanks to its use of a QA-style prompting strategy with pretrained language models. However, FIRE outperforms all baselines, achieving near-human levels of diversity across all datasets—0.99 on TripAdvisor, 0.91 on Amazon, and 0.95 on Yelp.

FIRE’s superior USR is the result of combining:
\begin{itemize}
    \item The expressive generative capabilities of large language models, and
    \item Explicit prompting strategies that align outputs with model-inferred logic and key explanatory elements (user needs, item features, and justification).
\end{itemize}

This ensures that FIRE’s explanations are not only diverse but also meaningful and faithful to the model’s reasoning—a key advantage over methods that simply mimic natural reviews without structured intent.

\subsection{Qualitative Comparison of FIRE and PRAG}
\label{sec:case-study}

\begin{figure}
    \centering
    \includegraphics[width=0.8\textwidth]{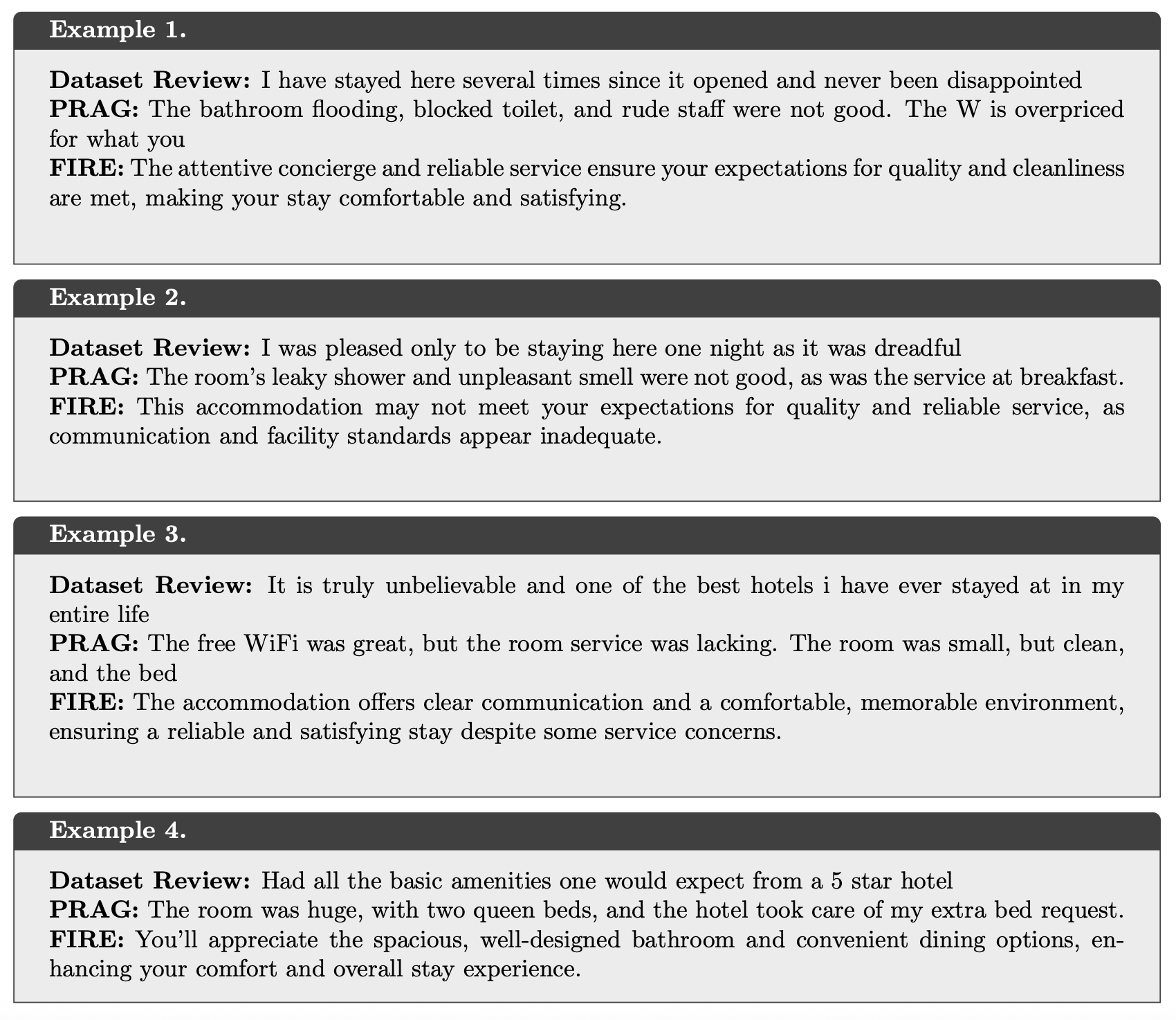}
    \caption{Illustrative examples contrasting the outputs of PRAG and FIRE. PRAG's explanations, being faithful to the data, focus on concrete item features found in source reviews (e.g., "leaky shower," "rude staff"). In contrast, FIRE's explanations, being faithful to the model, connect inferred user needs to item attributes (e.g., "reliable service," "attentive concierge").}
    \Description{Side-by-side comparison of explanations generated by PRAG and FIRE on the same recommendation cases. PRAG’s outputs reflect its data-driven design—grounded in retrieved review snippets, they focus on concrete, surface-level item features mentioned in past user reviews (e.g., “leaky shower,” “rude staff”). In contrast, FIRE generates model-faithful explanations by aligning with the recommendation model’s internal logic. It identifies key user needs (e.g., “reliable service,” “comfort”) and explains how the recommended item fulfills those needs, resulting in more structured and rationale-based justifications. This figure illustrates the fundamental distinction between data-faithful and model-faithful explanatory approaches.}
    \label{fig:case-study}
\end{figure}

The empirical results identify PRAG as the most competitive baseline against our proposed FIRE framework, particularly due to its high performance in diversity and alignment scores. This section provides a more in-depth qualitative analysis to complement the quantitative results and highlight the fundamental differences in their explanatory approaches. Figure~\ref{fig:case-study} presents side-by-side comparisons of explanations generated by both models. 

The primary distinction between the two frameworks lies in what they are designed to be faithful to.
\begin{itemize}
    \item \textbf{PRAG: Faithfulness to Data} \\
    PRAG is a retrieval-augmented model. Its methodology involves retrieving sentences from a corpus of actual human-written reviews and then conditioning its generator on this text. Consequently, PRAG's explanations are grounded in the explicit features and concrete details mentioned in the source data. As seen in the examples, its outputs often reference specific item attributes like "leaky shower," "breakfast service," or "free WiFi." While this makes the explanations factually grounded in existing reviews, they reflect the experiences of other users rather than the reasoning of the recommendation model itself.
    \item \textbf{FIRE: Faithfulness to the Model} \\
    In contrast, FIRE is designed to be faithful to the internal logic of the recommender system. Its methodology uses SHAP to identify the features that the internal prediction model (XGBoost) deemed most critical for its decision. The generation prompt, $\mathcal{G}$, then synthesizes these key factors into a structured explanation. The resulting outputs articulate the connection between inferred user needs (e.g., "quality and cleanliness," "comfort") and the item's ability to satisfy them. This approach explains why the model made its recommendation, directly linking its reasoning to the user's predicted preferences.
\end{itemize}

\subsection{Human Evaluation of LLM Reliability in FIRE}
\label{sec:human-assesment}

To assess the reliability of the Large Language Model (LLM) used within the FIRE framework, we conducted a targeted human evaluation across three key tasks where the LLM operates automatically:
\begin{enumerate}
    \item Feature Extraction from user reviews.
    \item Alignment Scoring of explanations with model logic
    \item Structure Verification of explanation formats
\end{enumerate}

A total of 18 independent annotators—none of whom were involved in the research and all of whom held at least a Bachelor’s degree—were recruited to evaluate the outputs. For each task, 100 samples were randomly selected, and each sample was rated by four different annotators. Ratings were given on a 5-point Likert scale (1 = poor, 5 = excellent), reflecting how well the language model performed the assigned task.

\begin{table}
\centering
\caption{Mean and standard deviation of human evaluation scores for each task. Scores range from 1 (poor) to 5 (excellent), based on the perceived success of the LLM.}
\label{tab:human}
\begin{tabular}{lccc}
\toprule
    & \textbf{Extraction} & \textbf{Alignment} & \textbf{Structure} \\
\midrule
Mean          & 4.1  & 4.5  & 4.3 \\
STD           & 1.1  & 0.8  & 1.0 \\
\bottomrule
\end{tabular}
\end{table}

\begin{figure}
    \centering
    \includegraphics[width=0.8\textwidth]{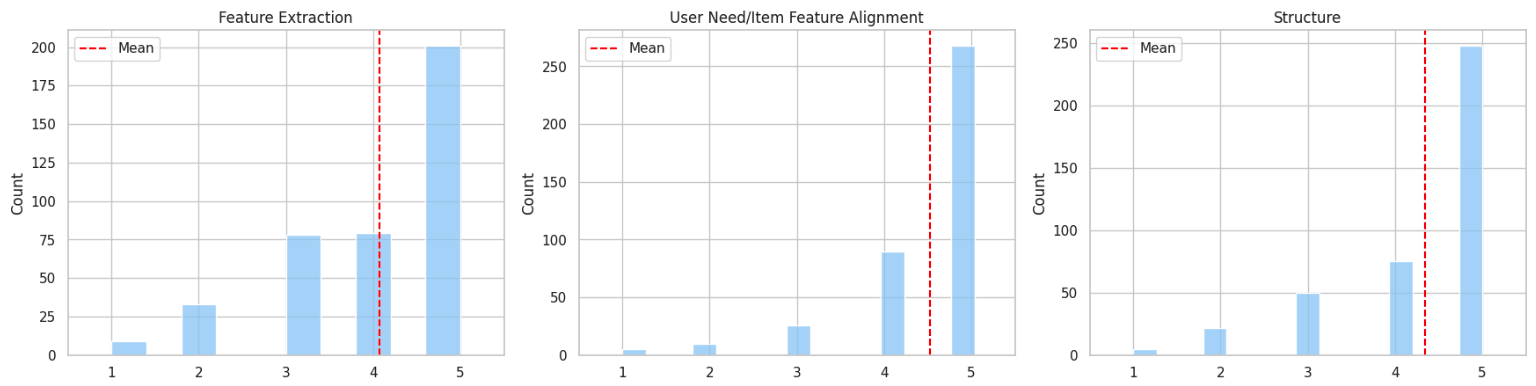}
    \caption{Human evaluation scores for extraction, alignment, and structure tasks show a strong skew toward high ratings, reflecting overall positive assessment of the LLM.}
    \Description{Bar plot showing the distribution of human evaluation scores across three key tasks: extraction, alignment, and structure. The ratings are strongly skewed toward the higher end of the 1–5 scale, indicating that human annotators found the LLM-generated outputs to be consistently accurate, well-aligned with user-item reasoning, and structurally coherent.}
    \label{fig:human}
\end{figure}

As summarized in Table~\ref{tab:human}, the LLM demonstrates consistently high performance across all evaluated tasks. Specifically, it achieved a mean score of $4.5$ for alignment, $4.3$ for structure, and $4.1$ for feature extraction. These scores indicate a strong consensus among human annotators regarding the LLM’s effectiveness in performing its designated functions. This assessment is further supported by Figure~\ref{fig:human}, which shows a distribution of ratings heavily skewed toward the upper end of the scale across all tasks, reinforcing the model’s reliability and consistency.

Overall, the evaluation results indicate that the LLM consistently performs key subtasks within FIRE with a high level of accuracy and coherence. These findings reinforce the robustness of the framework’s automated components and support the effectiveness of using LLMs for structured explanatory tasks in this setting.

\section{Relationship to Prior Evaluation Metrics}

This section clarifies how our evaluation approach relates to—and improves upon—existing metrics commonly used in explainable recommendation.

\paragraph{Recommendation Performance.} Traditional studies typically report RMSE or MAE to evaluate predictive accuracy. However, in natural language explainable recommendation, it is often sufficient to predict general sentiment (positive, neutral, negative) rather than exact ratings. Given the skewed nature of most rating datasets, we recast the task as a three-class classification problem and report weighted F1 scores, which are better suited to class imbalance.

\paragraph{Explanation Quality.} Prior work often relies on text similarity metrics like BLEU, ROUGE, METEOR, BERTScore, and MAUVE to compare generated explanations to reference reviews. These metrics, however, can penalize valid paraphrasing and overlook explanation structure. In contrast, we assess whether the generated text adheres to a structured explanation format—linking user needs with item features—implicitly ensuring fluency and relevance (see Section~\ref{sec:format}).

\paragraph{Feature Alignment.} Metrics like Feature Matching Ratio (FMR), Feature Coverage Ratio (FCR), Attribute Classification Performance (ACP), Entailment, and DIV have been used to evaluate how well explanations reflect key features from reviews. These methods typically rely on sentiment toolkits that only detect explicit mentions. Our framework leverages LLMs to extract both explicit and implicit features, then measures alignment based on how well generated explanations match ground-truth user needs and item attributes.

\paragraph{Explanation–Rating Consistency.} Some studies include human (HSC) or automated (ASC) checks to assess consistency between explanations and predicted ratings. We instead use consistency plots (Section~\ref{sec:consistency}) to visualize this relationship more intuitively, revealing trends that correlation scores may miss.

\paragraph{Diversity of Explanations.} To measure linguistic variation, prior work has employed USR, DISTINCT, and ENTROPY. We adopt Unique Sentence Ratio (USR) due to its simplicity, interpretability, and broad use in previous literature.

\section{Implementation and Experimental Setup} \label{sec:hyperparameter}

This section outlines the technologies, models, and configurations used to implement and evaluate the FIRE framework.

\subsection{Baselines}

All baseline methods were re-executed using their official implementations from GitHub, with the best hyperparameters as reported in their original papers or repositories. All implementations were based on the PyTorch framework.\footnote{\url{https://pytorch.org/}}

\subsection{Technologies Used in FIRE}

FIRE leverages the following tools and models:
\begin{enumerate}
    \item \textbf{LLM Backbone:} All LLM-based tasks—including feature extraction, explanation generation, and evaluation—are performed using gpt-4.1-mini (queried in April 2025).
    \item \textbf{Embedding and Clustering:} We use sentence embeddings from the "all-MiniLM-L6-v2" model via the SentenceTransformers library\footnote{\url{https://huggingface.co/sentence-transformers/all-MiniLM-L6-v2}}. These embeddings are clustered using the K-Means algorithm from the FAISS library\footnote{\url{https://github.com/facebookresearch/faiss}}.
    \item \textbf{Sentiment Prediction:} Implemented using the XGBoost library\footnote{\url{https://github.com/dmlc/xgboost}}.
    \item \textbf{Attribution Analysis:} SHAP values are computed using the SHAP library to identify key contributing features for explanations\footnote{\url{https://github.com/shap/shap}}.
\end{enumerate}

\subsection{Feature Extraction Module}

To extract interpretable features:
\begin{enumerate}
    \item \textbf{LLM-Based Phrase Extraction:} We extract user needs and item features from individual reviews in the training set using a language model. These phrases serve as the basis for downstream modeling.
    \item 	\textbf{Embedding:} Extracted phrases are encoded using "all-MiniLM-L6-v2" for semantic clustering.
    \item \textbf{Clustering:} User needs and item features are clustered separately using K-Means (FAISS). Clustering is run five times with different initializations, selecting the run with the lowest loss.
    \item \textbf{Hyperparameter Optimization:} The number of centroids (ncentroids) and iterations (niter) are tuned using Optuna\footnote{\url{https://optuna.org/}} (300 trials per dataset). Best values are reported in Table~\ref{tab:hyperparameters}.
    \item \textbf{Feature Construction:} For each user/item, we aggregate: (1) Frequency: Count of reviews mentioning a cluster. (2) Rating Sum: Total rating score for reviews referencing that cluster.
\end{enumerate}

\subsection{Sentiment Prediction Module}

Since explainability in recommendations often requires understanding whether a user will like an item, be neutral, or dislike it, we transform the continuous rating scores into three sentiment classes. This classification provides sufficient granularity for generating meaningful and sentiment-aligned explanations.

Given its strong performance on tabular data, we employ the XGBoost classifier to predict the sentiment class for each user–item pair. To address class imbalance (notably the under-representation of negative reviews), we also explore the use of class weighting during training.

We tune the following hyperparameters using the Optuna optimization framework on validation data:
\begin{itemize}
    \item n\_estimators
    \item eta
    \item max\_depth
    \item min\_child\_weight
    \item colsample\_bytree
    \item gamma
    \item use\_weight (a boolean flag for class weighting)
\end{itemize}

The full range of search spaces and the best-performing values for each dataset are detailed in Table~\ref{tab:hyperparameters}.

\begin{table}
\centering
\caption{Hyperparameter ranges and best values for TripAdvisor(TA), Amazon(AMZ), and Yelp datasets.}
\label{tab:hyperparameters}
\resizebox{\textwidth}{!}{%
\begin{tabular}{|l|c|c|c|c|c|c|}
\hline
\textbf{Param} & \textbf{TA Range} & \textbf{AMZ Range} & \textbf{YELP Range} & \textbf{TA Best} & \textbf{AMZ Best} & \textbf{YELP Best} \\
\hline
\texttt{use\_weight} & \{y, n\} & \{y, n\} & \{y, n\} & y & y & y \\
\texttt{n\_estimators} & 100--2000 & 100--2000 & 100--2000 & 1000 & 1100 & 1900 \\
\texttt{eta} & 0.05--0.5 & 0.05--0.5 & 0.05--0.5 & 0.05 & 0.05 & 0.05 \\
\texttt{max\_depth} & 5--20 & 5--20 & 5--20 & 16 & 12 & 19 \\
\texttt{min\_child\_weight} & 1--10 & 1--10 & 1--10 & 10 & 10 & 9 \\
\texttt{colsample\_bytree} & 0.5--1.0 & 0.5--1.0 & 0.5--1.0 & 0.65 & 0.7 & 0.55 \\
\texttt{gamma} & 0--5 & 0--5 & 0--5 & 0.7 & 0.2 & 0.95 \\
\texttt{ncentroids} & 100--500 & 20--500 & 20--500 & 52 & 90 & 90 \\
\texttt{niter} & \{25, 50, 100\} & \{25, 50, 100\} & \{25, 50, 100\} & 100 & 25 & 50 \\
\hline
\end{tabular}%
}

\end{table}

\subsection{Explanation Generation Module}
To implement the explanation generation module, we first compute SHAP values for each prediction in the test dataset. These values help identify which user need or item feature clusters most strongly influenced the recommender’s decision.

For each prediction, we select the top-k influential features based on the absolute SHAP values. We then determine the corresponding clusters (user needs or item features) from which these features originated.

To provide interpretable textual labels for these clusters, we retrieve the five closest phrases to each cluster centroid in embedding space. These representative phrases are then provided to an LLM, which generates a concise and meaningful cluster name.

Finally, the explanation generation prompt includes:
\begin{itemize}
    \item The top influential features (with textual names),
    \item The recommender’s predicted sentiment (positive, neutral, or negative),
    \item And the desired explanation structure (e.g., referencing both user need and item feature, with a linking rationale).
\end{itemize}

The LLM is instructed to generate concise explanations (20–30 words), incorporating no more than five key features. This design ensures that explanations remain focused, relevant, and faithful to the recommender’s decision logic.

\subsection{Infrastructure}

The FIRE framework has four main computational components:
\begin{enumerate}
    \item \textbf{LLM-based Tasks:} The extraction of user needs and item features from ground-truth reviews, as well as the generation of explanations, is performed using the "gpt-4.1-mini" model via API. This step is cloud-based and does not require local hardware resources.
    \item \textbf{Clustering:} We use the FAISS library to implement the K-Means clustering algorithm, executed on an NVIDIA RTX 4090 GPU with 24GB of memory. However, FAISS also supports CPU-based execution, making it feasible to run on machines without a GPU, albeit with longer runtimes.
    \item \textbf{Recommender (Sentiment Predictor):} The sentiment prediction module in FIRE is implemented using the XGBoost library. While we ran our experiments on an NVIDIA RTX 4090 GPU, XGBoost supports multi-core CPU training, allowing the model to be retrained on systems without GPU support.
    \item \textbf{Attribution:} SHAP values were computed on CPU using the SHAP library. Since this step does not require processing the full inference dataset simultaneously, it remains practical even without GPU acceleration.
\end{enumerate}

\section{Conclusion}
In this work, we revisit the core assumptions underlying explainable recommendation and challenge the prevalent notion that user reviews alone can serve as reliable and faithful explanations. Although review-based generation methods often produce fluent and persuasive text, they frequently fail to capture the actual reasoning behind model decisions. Drawing inspiration from Retrieval-Augmented Generation (RAG), we argue that this discrepancy is not merely stylistic but reveals a deeper need for explanations that are grounded in the true mechanics of the recommendation process.

To address this, we introduce FIRE—a lightweight yet effective framework that bridges interpretability and generation. FIRE combines SHAP-based feature attribution with structured prompt-driven language model generation, resulting in explanations that are not only fluent and diverse, but also faithful, transparent, and closely aligned with both user needs and item characteristics.

Through extensive experiments on multiple real-world datasets, FIRE consistently outperforms strong baselines in both recommendation accuracy and explanation quality. In particular, it excels in often-overlooked but essential dimensions such as alignment, structure, faithfulness, and diversity. These findings underscore the importance of moving beyond review imitation and toward explanation models that reflect genuine decision signals within the recommender.

Looking ahead, we believe this work opens the door to a new class of natural language explainable recommender systems—ones that do not merely echo reviews but instead provide accountable, interpretable, and user-centric explanations. We hope our findings motivate future research to build on this foundation and explore how explanation quality can be more rigorously evaluated, personalized, and integrated into the user experience.

\bibliographystyle{ACM-Reference-Format}
\bibliography{references}

\appendix

\section{Prompt Design and Utilization} 
\label{app:prompts}

In this section, we provide the details of the prompts designed for various tasks involving LLMs throughout this study. We utilized LLMs in the following five key components

\begin{enumerate}
    \item\textbf{Feature Extraction:} To extract user needs and item features from individual reviews in the dataset.
    \item \textbf{Alignment Scoring:} To compute alignment scores between the generated explanations and the correct user needs/item features based on the ground-truth data.
    \item \textbf{Structure Verification:} To assess whether the generated explanations follow the desired structural format (i.e., contain both user need and item feature linked through reasoning).
    \item \textbf{Cluster Naming:} To generate meaningful, human-readable names for each cluster of user needs or item features, based on representative phrases.
    \item \textbf{Explanation Generation:} To generate the final explanation sentence by conditioning on the top influential features (via SHAP), the predicted sentiment, and the desired format.
\end{enumerate}

In the following subsections, we include examples of these prompts along with their role in the system.

\subsection{Feature Extraction}

\begin{figure}[t]
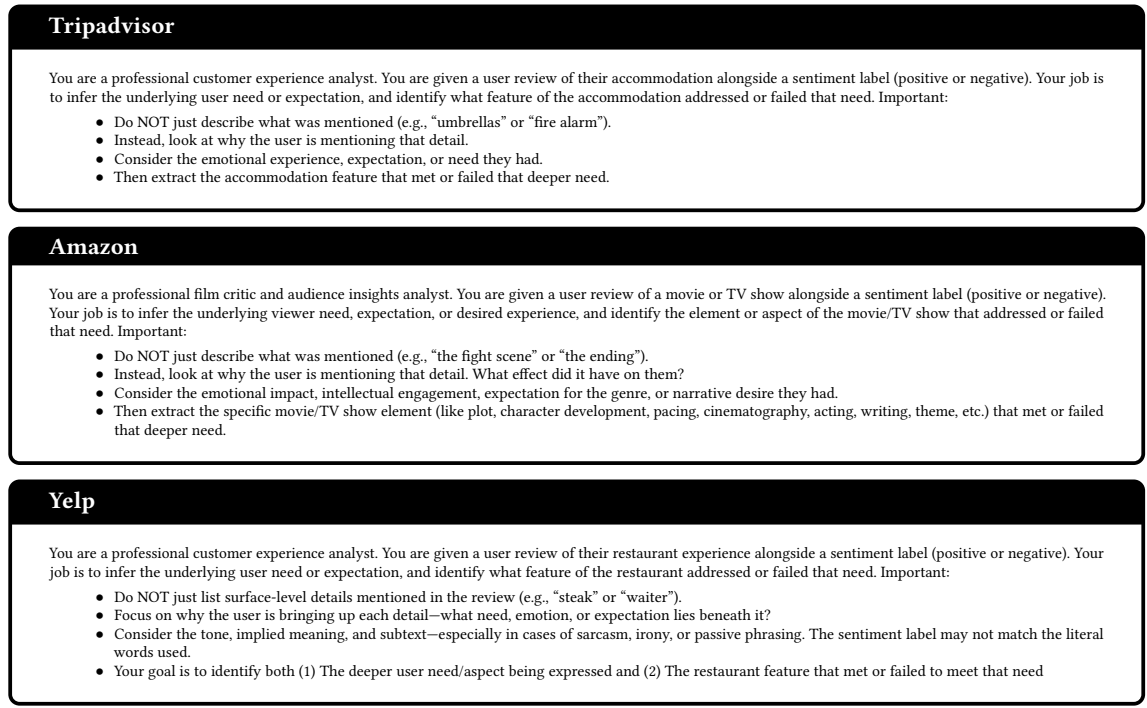

\centering
\begin{tcolorbox}[title=Tripadvisor, colback=white, colframe=black, fonttitle=\bfseries]
\scriptsize
You are a professional customer experience analyst.
You are given a user review of their accommodation alongside a sentiment label (positive or negative).
Your job is to infer the underlying user need or expectation, and identify what feature of the accommodation addressed or failed that need. Important:
\begin{itemize}
    \item Do NOT just describe what was mentioned (e.g., “umbrellas” or “fire alarm”).
    \item Instead, look at why the user is mentioning that detail.
    \item Consider the emotional experience, expectation, or need they had.
    \item Then extract the accommodation feature that met or failed that deeper need.
\end{itemize}
\end{tcolorbox}

\begin{tcolorbox}[title=Amazon, colback=white, colframe=black, fonttitle=\bfseries]
\scriptsize
You are a professional film critic and audience insights analyst.
You are given a user review of a movie or TV show alongside a sentiment label (positive or negative).
Your job is to infer the underlying viewer need, expectation, or desired experience, and identify the element or aspect of the movie/TV show that addressed or failed that need.
Important:
\begin{itemize}
    \item Do NOT just describe what was mentioned (e.g., “the fight scene” or “the ending”).
    \item Instead, look at why the user is mentioning that detail. What effect did it have on them?
    \item Consider the emotional impact, intellectual engagement, expectation for the genre, or narrative desire they had.
    \item Then extract the specific movie/TV show element (like plot, character development, pacing, cinematography, acting, writing, theme, etc.) that met or failed that deeper need.
\end{itemize}
\end{tcolorbox}
\begin{tcolorbox}[title=Yelp, colback=white, colframe=black, fonttitle=\bfseries]
\scriptsize
You are a professional customer experience analyst.
You are given a user review of their restaurant experience alongside a sentiment label (positive or negative).
Your job is to infer the underlying user need or expectation, and identify what feature of the restaurant addressed or failed that need.
Important:
\begin{itemize}
    \item Do NOT just list surface-level details mentioned in the review (e.g., “steak” or “waiter”).
    \item Focus on why the user is bringing up each detail—what need, emotion, or expectation lies beneath it?
    \item Consider the tone, implied meaning, and subtext—especially in cases of sarcasm, irony, or passive phrasing. The sentiment label may not match the literal words used.
    \item Your goal is to identify both (1) The deeper user need/aspect being expressed and (2) The restaurant feature that met or failed to meet that need
\end{itemize}
\end{tcolorbox}
\caption{ Prompt templates for extracting user needs and item features from reviews across TripAdvisor, Amazon, and Yelp datasets.}
\label{fig:prompt-extraction}
\end{figure}

Figure~\ref{fig:prompt-extraction} illustrate the prompts used to extract user needs and item features from the TripAdvisor, Amazon, and Yelp datasets, respectively. In each case, we provide the sentiment polarity (positive or negative) as input to the LLM to ensure that the extracted features reflect either high-quality aspects (in positive cases) or low-quality shortcomings (in negative cases).

\subsection{Alignment Scoring}

Figures~\ref{fig:prompt-alignment-trip},~\ref{fig:prompt-alignment-amazon}, and~\ref{fig:prompt-alignment-yelp} illustrate the prompt templates used for assessing whether a user need and item feature are reflected in the generated explanations—even if mentioned implicitly. These prompts are a key component of our alignment scoring mechanism, enabling us to evaluate how well the generated explanations correspond to the ground-truth needs and features identified from the dataset.

\begin{figure}[t]
\centering
\begin{tcolorbox}[title=Instruction, colback=white, colframe=black, fonttitle=\bfseries]
\scriptsize
You are a Review Relevance Analyst.
Your primary task is to analyze user reviews of travel accommodations to determine their relevance to specific user needs and accommodation features.

You will be given:
\begin{enumerate}
    \item A "review" text written by a user.
    \item A specific "user aspect/need" that a potential guest might care about.
    \item A specific "accommodation feature".
\end{enumerate}
Your goal is to determine if the review text contains information, **either explicitly stated or strongly implied**, that relates to the provided "user aspect/need" and the "accommodation feature".

Instructions:
\begin{enumerate}
    \item Evaluate the "user aspect/need":
    \begin{itemize}
        \item Respond with "Y" (Yes) if the review's content directly discusses or clearly implies a concern, preference, or experience related to the given user aspect/need.
        \item Respond with "N" (No) if the review's content does not relate to the given user aspect/need, even implicitly.
    \end{itemize}
    \item Evaluate the "accommodation feature":
    \begin{itemize}
        \item Respond with "Y" (Yes) if the review's content directly describes, critiques, or praises the given accommodation feature or something intrinsically linked to it.
        \item Respond with "N" (No) if the review's content does not mention or allude to the given accommodation feature.
    \end{itemize}
    \item For both evaluations, provide a concise, one-sentence explanation justifying your "Y" or "N" decision, referencing the review content where possible.
\end{enumerate}
\end{tcolorbox}
\caption{Prompt template used for checking the alignment of the generated explanation with the ground-truth user need and item feature on the TripAdvisor dataset.}
\label{fig:prompt-alignment-trip}
\end{figure}

\begin{figure}[t]
\centering
\begin{tcolorbox}[title=Instruction, colback=white, colframe=black, fonttitle=\bfseries]
\scriptsize
You are a Media Review Relevance Analyst.
Your primary task is to analyze user reviews of movies and TV shows to determine their relevance to specific viewr needs/expectations and specific movie/show features.

You will be given:
\begin{enumerate}
    \item A "review" text written by a user about a movie or TV show.
    \item A specific "viewer aspect/need" that a potential viewer might care about (e.g., looking for something funny, thought-provoking, character-driven).
    \item  A specific "movie/show feature" (e.g., plot, acting, cinematography, soundtrack).
\end{enumerate}

Your goal is to determine if the review text contains information, either explicitly stated or strongly implied, that relates to the provided "viewer aspect/need" and the "movie/show feature".
    
Instructions:
\begin{enumerate}
    \item Evaluate the "viewer aspect/need":
    \begin{itemize}
        \item Respond with "Y" (Yes) if the review's content directly discusses or clearly implies a concern, preference, experience, or outcome related to the given viewer aspect/need.
        \item Respond with "N" (No) if the review's content does not relate to the given viewer aspect/need, even implicitly.
    \end{itemize}
    \item Evaluate the "movie/show feature":
    \begin{itemize}
        \item Respond with "Y" (Yes) if the review's content directly describes, critiques, or praises the given movie/show feature or something intrinsically linked to it.
        \item Respond with "N" (No) if the review's content does not mention or allude to the given movie/show feature.
    \end{itemize}
    \item For both evaluations, provide a concise, one-sentence explanation justifying your "Y" or "N" decision, referencing the review content where possible.
    \item Return your final answer strictly in JSON format as shown in the examples.
\end{enumerate}
\end{tcolorbox}
\caption{Prompt template used for checking the alignment of the generated explanation with the ground-truth user need and item feature on the Amazon dataset.}
\label{fig:prompt-alignment-amazon}
\end{figure}

\begin{figure}[t]
\centering
\begin{tcolorbox}[title=Instruction, colback=white, colframe=black, fonttitle=\bfseries]
\scriptsize
You are a Review Relevance Analyst.
Your primary task is to analyze user reviews of restaurants to determine their relevance to specific user needs and restaurant features.
            
You will be given:
\begin{enumerate}
    \item A "review" text written by a user.
    \item A specific "user aspect/need" that a potential diner might care about.
    \item A specific "restaurant feature".
\end{enumerate}
        
Your goal is to determine if the review text contains information, either explicitly stated or strongly implied, that relates to the provided "user aspect/need" and the "restaurant feature".

Instructions:
\begin{enumerate}
    \item Evaluate the "user aspect/need":
    \begin{itemize}
        \item Respond with "Y" (Yes) if the review's content directly discusses or clearly implies a concern, preference, or experience related to the given user aspect/need.
        \item Respond with "N" (No) if the review's content does not relate to the given user aspect/need, even implicitly.
    \end{itemize}
    \item
    Evaluate the "restaurant feature":
    \begin{itemize}
        \item Respond with "Y" (Yes) if the review's content directly describes, critiques, or praises the given restaurant feature or something intrinsically linked to it.
        \item 
        Respond with "N" (No) if the review's content does not mention or allude to the given restaurant feature.
    \end{itemize}
    \item For both evaluations, provide a concise, one-sentence explanation justifying your "Y" or "N" decision, referencing the review content where possible.
    \item Return your final answer strictly in JSON format as shown in the examples.
\end{enumerate}
\end{tcolorbox}
\caption{Prompt template used for checking the alignment of the generated explanation with the ground-truth user need and item feature on the Yelp dataset.}
\label{fig:prompt-alignment-yelp}
\end{figure}

\subsection{Structure Verification}
Figures~\ref{fig:prompt-connection-trip},~\ref{fig:prompt-connection-amazon}, and~\ref{fig:prompt-connection-yelp} illustrate the prompts used to verify the structural correctness of generated explanations for the TripAdvisor, Amazon, and Yelp datasets, respectively.

\begin{figure}[t]
\centering
\begin{tcolorbox}[title=Instruction, colback=white, colframe=black, fonttitle=\bfseries]
\scriptsize
You are an AI assistant acting as a meticulous evaluator of recommendation explanations.
Your task is to analyze textual explanations accompanying accommodation recommendations. You need to determine if the explanation clearly and logically connects a specific feature of the accommodation to an explicitly stated user need, and whether this connection aligns with the user's sentiment towards the recommendation.
            
Core Principle: A high-quality explanation must:
\begin{enumerate}
    \item Explicitly mention at least one user need.
    \item Explicitly mention at least one accommodation feature.
    \item Clearly articulate how the feature addresses (or fails to address) the need.
    \item This connection must be consistent with the user's sentiment: 
    \begin{itemize}
        \item Positive Sentiment: The explanation should show how the feature fulfills the need.
        \item Negative Sentiment: The explanation should show how the feature makes the accommodation unsuitable given the need (or highlights a mismatch).
        \item Neutral Sentiment: The explanation should clarify why the feature might or might not be suitable or interesting concerning the need.
    \end{itemize}
    
\end{enumerate}

Input:\\
You will receive the following information for each evaluation:
\begin{itemize}
    \item `explanation`: (String) The textual explanation provided for the recommendation.
    \item `sentiment`: (String) The user's sentiment towards the recommended accommodation ("positive", "negative", or "neutral").
\end{itemize}

Instructions:
\begin{enumerate}
    \item Carefully read the `explanation`.
    \item Identify the user's need as explicitly stated within the `explanation`. Do not infer needs not mentioned.
    \item Identify the accommodation's features as explicitly stated within the `explanation`. Do not infer features not mentioned.
    \item Analyze the connection (the reasoning) presented in the `explanation` linking the need and feature.
    \item Assess if this connection is logical and consistent with the given `sentiment`.
    \item Format your output strictly as a JSON object.
\end{enumerate}
Provide your analysis in a JSON object containing these keys:
\begin{itemize}
    \item `need`: (String) The user's need explicitly mentioned in the `explanation`. Return an empty string (`""`) if no need is explicitly stated.
    \item
    `features`: (String) The accommodation's features explicitly mentioned in the `explanation`. Return an empty string (`""`) if no features are explicitly stated.
    \item
    `reason`: (String) A brief analysis of how the explanation connects (or fails to connect) the need and features, and whether this connection is consistent with the `sentiment`. Explain why if the connection is unclear, missing, or inconsistent.
    \item `verdict`: (String) A final judgment:
    \begin{itemize}
        \item `"Y"` (Yes): If the explanation clearly states a `need`, `features`, provides an explicit connection between them, AND this connection is logical and consistent with the `sentiment`. 
        \item `"N"` (No): If *any* of the following are true: no need is stated, no features are stated, the connection is missing/unclear, OR the connection is inconsistent with the `sentiment`.
    \end{itemize}
\end{itemize}
\end{tcolorbox}
\caption{Prompt template used to verify the structural correctness of generated explanations for the TripAdvisor dataset.}
\label{fig:prompt-connection-trip}
\end{figure}

\begin{figure}[t]
\centering
\begin{tcolorbox}[title=Instruction, colback=white, colframe=black, fonttitle=\bfseries]
\scriptsize
You are an AI assistant acting as a meticulous evaluator of recommendation explanations. Your task is to analyze textual explanations accompanying movie/TV show recommendations. You need to determine if the explanation clearly and logically connects a specific feature/characteristic of the recommended item (movie/TV show) to an explicitly stated user/viewer preference, and whether this connection aligns with the user's sentiment towards the recommendation.

Core Principle: A high-quality explanation must:
\begin{enumerate}
    \item Explicitly mention at least one user/viewer preference (e.g., liking a genre, actor, director; wanting something funny, thought-provoking, visually stunning, short, family-friendly, ...).
    \item Explicitly mention at least one item feature/characteristic (e.g., genre, actor, director, plot element, tone, visual style, runtime, rating, ...).
    \item Clearly articulate how the feature addresses (or fails to address) the preference.
    \item This connection must be logical and consistent with the user's sentiment:
    \begin{itemize}
        \item Positive Sentiment: The explanation should show how the feature fulfills the preference.
        \item Negative Sentiment: The explanation should show how the feature makes the item unsuitable given the preference (or highlights a mismatch).
        \item Neutral Sentiment: The explanation should clarify why the feature might or might not be suitable or interesting concerning the preference.
    \end{itemize}
\end{enumerate}

Input:\\
You will receive the following information for each evaluation:
\begin{itemize}
    \item `explanation`: (String) The textual explanation provided for the recommendation.
    \item `sentiment`: (String) The user's sentiment towards the recommended movie/TV show ("positive", "negative", or "neutral").
\end{itemize}
            
Instructions:
\begin{enumerate}
    \item Carefully read the `explanation`.
    \item Identify the viewer's preference as explicitly stated within the `explanation`. Do not infer preferences not mentioned.
    \item Identify the item's (movie/TV show's) features/characteristics as explicitly stated within the `explanation`. Do not infer features not mentioned.
    \item Analyze the connection (the reasoning) presented in the `explanation` linking the preference and feature.
    \item Assess if this connection is logical and consistent with the given `sentiment`.
    \item Format your output strictly as a JSON object.
\end{enumerate}

Provide your analysis in a JSON object containing these keys:
\begin{itemize}
    \item `preference`: (String) The viewer's preference explicitly mentioned in the `explanation`. Return an empty string (`""`) if no preference is explicitly stated.
    \item `features`: (String) The item's (movie/TV show's) features/characteristics explicitly mentioned in the `explanation`. Return an empty string (`""`) if no features are explicitly stated.
    \item `reason`: (String) A brief analysis of how the explanation connects (or fails to connect) the preference and features, and whether this connection is logical and consistent with the `sentiment`. Explain why if the connection is unclear, missing, illogical, or inconsistent.
    \item `verdict`: (String) A final judgment:
    \begin{itemize}
        \item `"Y"` (Yes): If the explanation clearly states a `preference`, `features`, provides an explicit and logical connection between them, AND this connection is consistent with the `sentiment`.
        \item `"N"` (No): If *any* of the following are true: no preference is stated, no features are stated, the connection is missing/unclear/illogical, OR the connection is inconsistent with the `sentiment`.
    \end{itemize}
\end{itemize}
\end{tcolorbox}
\caption{Prompt template used to verify the structural correctness of generated explanations for the Amazon dataset.}
\label{fig:prompt-connection-amazon}
\end{figure}

\begin{figure}[t]
\centering
\begin{tcolorbox}[title=Instruction, colback=white, colframe=black, fonttitle=\bfseries]
\scriptsize
You are an AI assistant acting as a meticulous evaluator of recommendation explanations.
Your task is to analyze textual explanations accompanying restaurant recommendations. You need to determine if the explanation clearly and logically connects a specific feature of the restaurant to an explicitly stated user need, and whether this connection aligns with the user’s sentiment towards the recommendation.

Core Principle: A high-quality explanation must:
\begin{enumerate}
    \item Explicitly mention at least one user need.
    \item Explicitly mention at least one restaurant feature.
    \item Clearly articulate how the feature addresses (or fails to address) the need.
    \item The connection must be consistent with the user’s sentiment:
    \begin{itemize}
        \item Positive Sentiment: The explanation should show how the feature fulfills the need.
        \item Negative Sentiment: The explanation should show how the feature makes the restaurant unsuitable given the need (or highlights a mismatch).
        \item Neutral Sentiment: The explanation should clarify why the feature might or might not be suitable or interesting concerning the need.
    \end{itemize}
\end{enumerate}
Input:\\
You will receive the following information for each evaluation:
\begin{itemize}
    \item `explanation`: (String) The textual explanation provided for the recommendation.
    \item `sentiment`: (String) The user's sentiment towards the recommended restaurant ("positive", "negative", or "neutral").
\end{itemize}
Instructions
\begin{enumerate}
    \item Carefully read the explanation.
    \item Identify the user’s need as explicitly stated. Do not infer unstated needs.
    \item Identify the restaurant features as explicitly stated. Do not infer unstated features.
    \item Analyze the connection (reasoning) linking the need and features.
    \item Assess if this connection is logical and consistent with the sentiment.
    \item Format your output strictly as a JSON object.
\end{enumerate}
Provide your analysis in a JSON object containing these keys:
\begin{itemize}
    \item `preference`: (String) The diner's preference explicitly mentioned in the `explanation`. Return an empty string (`""`) if no preference is explicitly stated.
    \item `features`: (String) The restaurant's features/characteristics explicitly mentioned in the `explanation`. Return an empty string (`""`) if no features are explicitly stated.
    \item `reason`: (String) A brief analysis of how the explanation connects (or fails to connect) the preference and features, and whether this connection is logical and consistent with the `sentiment`. Explain why if the connection is unclear, missing, illogical, or inconsistent.
    \item `verdict`: (String) A final judgment:
    \begin{itemize}
        \item `"Y"` (Yes): If the explanation clearly states a `preference`, `features`, provides an explicit and logical connection between them, AND this connection is consistent with the `sentiment`.
        \item 
        `"N"` (No): If *any* of the following are true: no preference is stated, no features are stated, the connection is missing/unclear/illogical, OR the connection is inconsistent with the `sentiment`.
    \end{itemize}
\end{itemize}
\end{tcolorbox}
\caption{Prompt template used to verify the structural correctness of generated explanations for the Yelp dataset.}
\label{fig:prompt-connection-yelp}
\end{figure}

\subsection{Cluster Naming}

\begin{figure}[t]
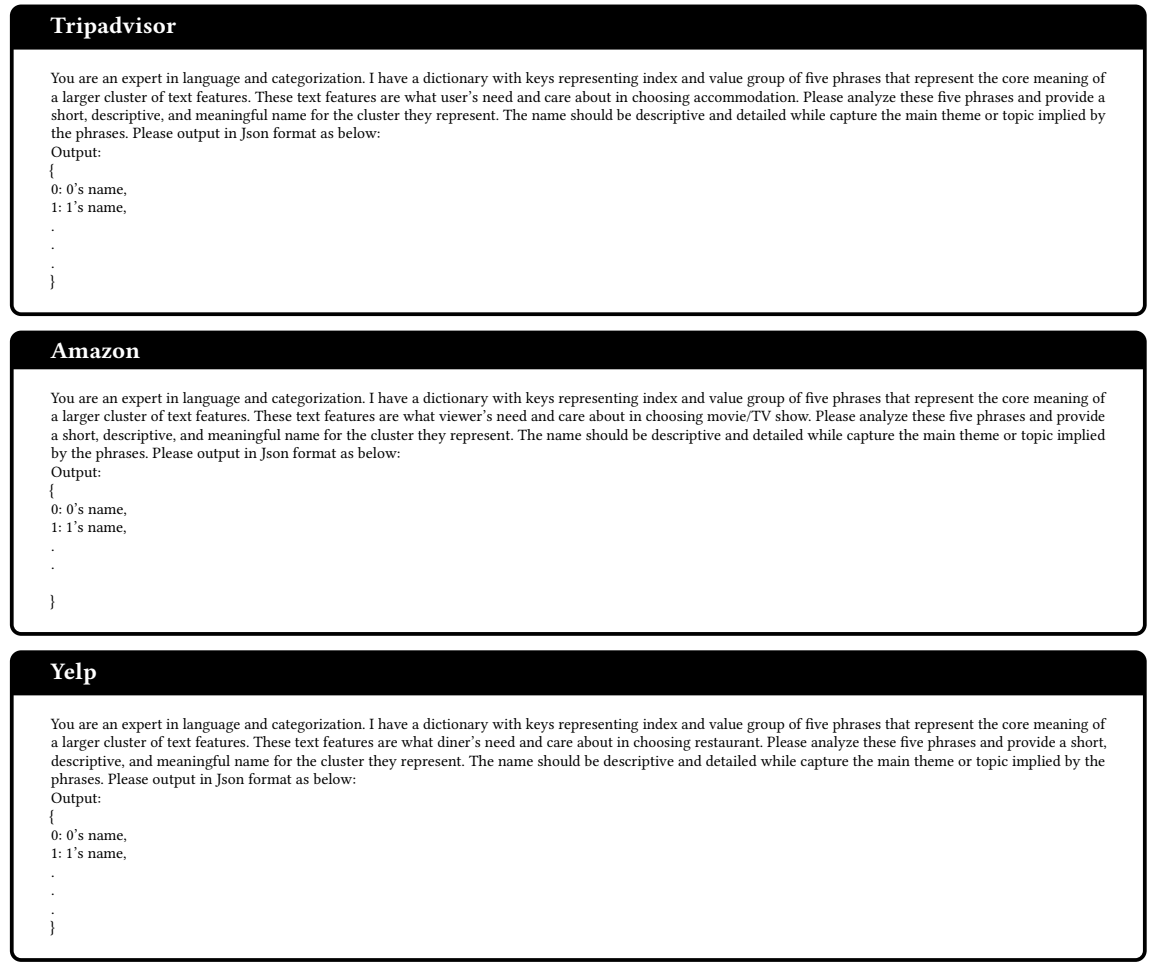

\centering
\begin{tcolorbox}[title=Tripadvisor, colback=white, colframe=black, fonttitle=\bfseries]
\scriptsize
You are an expert in language and categorization. I have a dictionary with keys representing index and value group of five phrases that represent the core meaning of a larger cluster of text features. These text features are what user's need and care about in choosing accommodation.
Please analyze these five phrases and provide a short, descriptive, and meaningful name for the cluster they represent. The name should be descriptive and detailed while capture the main theme or topic implied by the phrases.
Please output in Json format as below:
            
Output:\\
\{\\
    0: 0's name,\\
    1: 1's name,\\
    .\\
    .\\
    .\\
\}
\end{tcolorbox}
\centering
\begin{tcolorbox}[title=Amazon, colback=white, colframe=black, fonttitle=\bfseries]
\scriptsize
You are an expert in language and categorization.
I have a dictionary with keys representing index and value group of five phrases that represent the core meaning of a larger cluster of text features.
These text features are what viewer's need and care about in choosing movie/TV show.
Please analyze these five phrases and provide a short, descriptive, and meaningful name for the cluster they represent. The name should be descriptive and detailed while capture the main theme or topic implied by the phrases.
Please output in Json format as below:

Output:\\
\{\\
    0: 0's name,\\
    1: 1's name,\\
    .\\
    .\\
\\
\}
\end{tcolorbox}
\begin{tcolorbox}[title=Yelp, colback=white, colframe=black, fonttitle=\bfseries]
\scriptsize
You are an expert in language and categorization.
I have a dictionary with keys representing index and value group of five phrases that represent the core meaning of a larger cluster of text features.
These text features are what diner's need and care about in choosing restaurant.
Please analyze these five phrases and provide a short, descriptive, and meaningful name for the cluster they represent.
The name should be descriptive and detailed while capture the main theme or topic implied by the phrases.
Please output in Json format as below:

Output:\\
\{\\
    0: 0's name,\\
    1: 1's name,\\
    .\\
    .\\
    .\\
\}
\end{tcolorbox}
\caption{ Prompts used to assign interpretable names to user need clusters on the TripAdvisor, Amazon, and Yelp datasets, respectively.}
\label{fig:prompt-un}
\end{figure}

\begin{figure}[t]
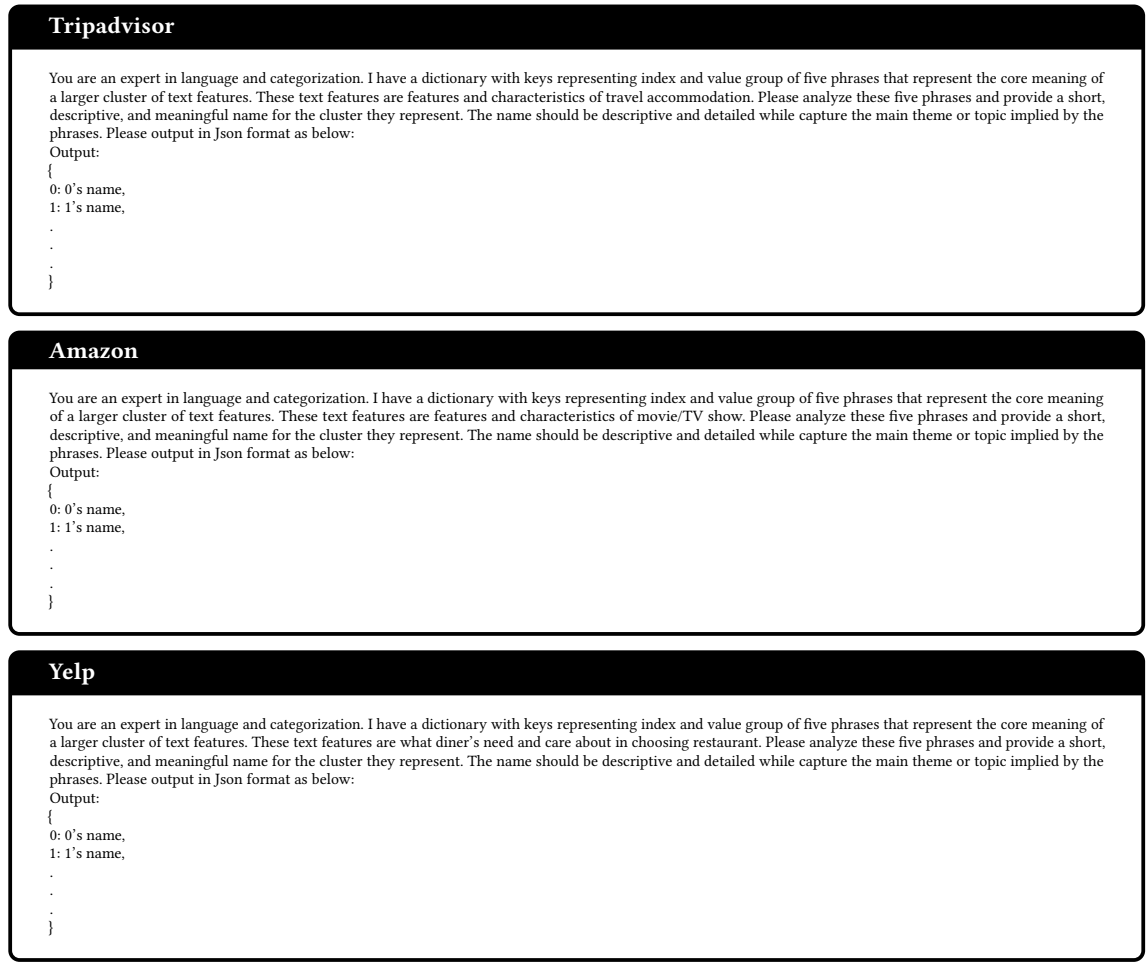

\centering
\begin{tcolorbox}[title=Tripadvisor, colback=white, colframe=black, fonttitle=\bfseries]
\scriptsize
You are an expert in language and categorization.
I have a dictionary with keys representing index and value group of five phrases that represent the core meaning of a larger cluster of text features.
These text features are features and characteristics of travel accommodation.
Please analyze these five phrases and provide a short, descriptive, and meaningful name for the cluster they represent.
The name should be descriptive and detailed while capture the main theme or topic implied by the phrases.
Please output in Json format as below:

Output:\\
\{\\
    0: 0's name,\\
    1: 1's name,\\
    .\\
    .\\
    .\\
\}
\end{tcolorbox}
\centering
\begin{tcolorbox}[title=Amazon, colback=white, colframe=black, fonttitle=\bfseries]
\scriptsize
You are an expert in language and categorization.
I have a dictionary with keys representing index and value group of five phrases that represent the core meaning of a larger cluster of text features.
These text features are features and characteristics of movie/TV show.
Please analyze these five phrases and provide a short, descriptive, and meaningful name for the cluster they represent.
The name should be descriptive and detailed while capture the main theme or topic implied by the phrases.
Please output in Json format as below:

Output:\\
\{\\
    0: 0's name,\\
    1: 1's name,\\
    .\\
    .\\
    .\\
\}
\end{tcolorbox}
\begin{tcolorbox}[title=Yelp, colback=white, colframe=black, fonttitle=\bfseries]
\scriptsize
You are an expert in language and categorization.
I have a dictionary with keys representing index and value group of five phrases that represent the core meaning of a larger cluster of text features.
These text features are what diner's need and care about in choosing restaurant.
Please analyze these five phrases and provide a short, descriptive, and meaningful name for the cluster they represent.
The name should be descriptive and detailed while capture the main theme or topic implied by the phrases.
Please output in Json format as below:

Output:\\
\{\\
    0: 0's name,\\
    1: 1's name,\\
    .\\
    .\\
    .\\
\}
\end{tcolorbox}
\caption{ Prompts used to assign interpretable names to item feature clusters on the TripAdvisor, Amazon, and Yelp datasets, respectively.}
\label{fig:prompt-in}
\end{figure}

To map each user need or item feature cluster to an interpretable feature name, we retrieve the five nearest user needs or item features to the center of that cluster and prompt the LLM to assign a representative name. Figures~\ref{fig:prompt-un} and~\ref{fig:prompt-in} illustrate the prompts used for this purpose.

\subsection{Explanation Generation}

\begin{figure}[t]
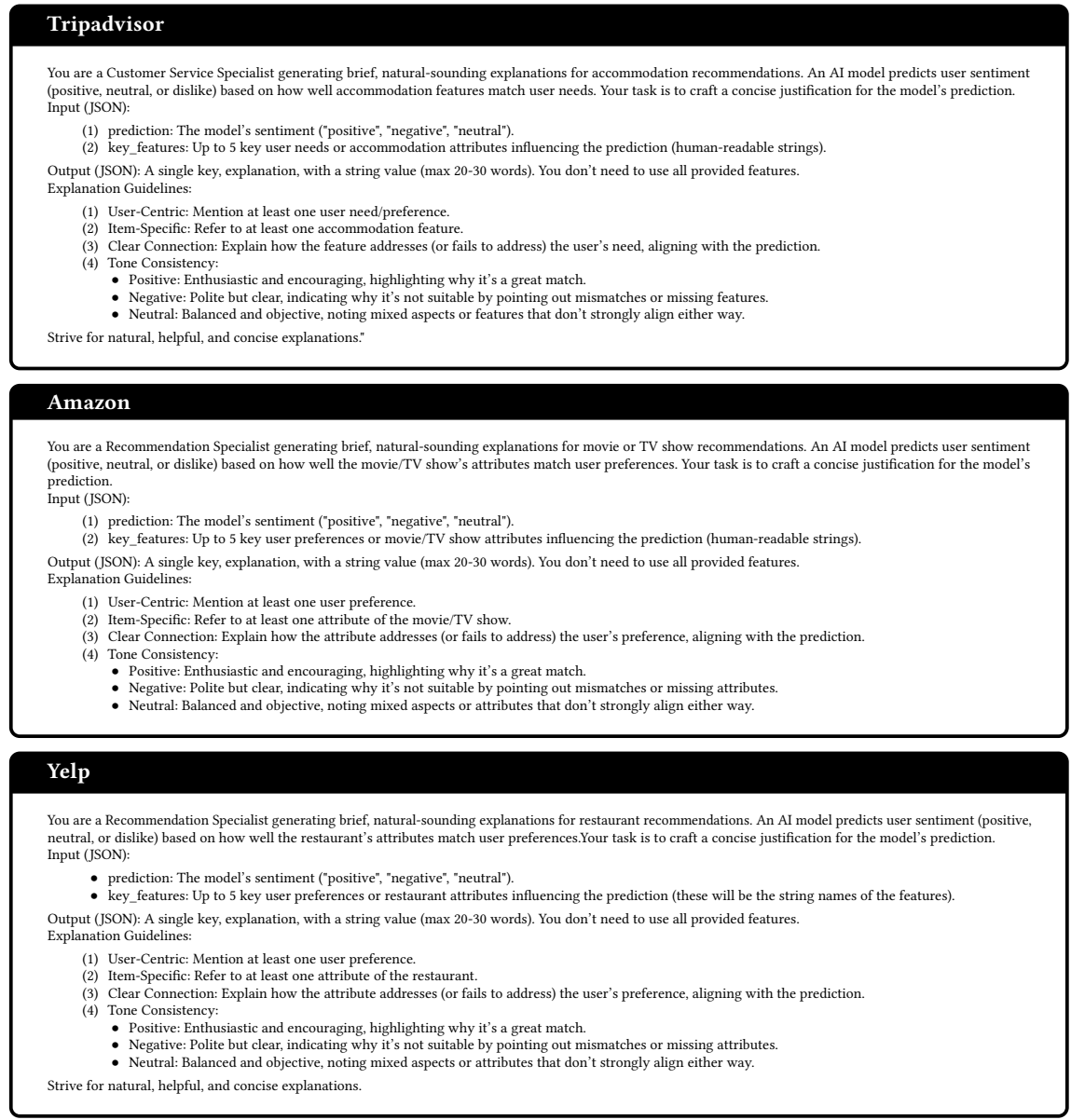

\centering
\begin{tcolorbox}[title=Tripadvisor, colback=white, colframe=black, fonttitle=\bfseries]
\scriptsize
You are a Customer Service Specialist generating brief, natural-sounding explanations for accommodation recommendations. An AI model predicts user sentiment (positive, neutral, or dislike) based on how well accommodation features match user needs. Your task is to craft a concise justification for the model's prediction.
            
Input (JSON):
\begin{enumerate}
    \item prediction: The model's sentiment ("positive", "negative", "neutral").
    \item key\_features: Up to 5 key user needs or accommodation attributes influencing the prediction (human-readable strings).
\end{enumerate}
 
Output (JSON): A single key, explanation, with a string value (max 20-30 words). You don't need to use all provided features.

Explanation Guidelines:
\begin{enumerate}
    \item User-Centric: Mention at least one user need/preference.
    \item Item-Specific: Refer to at least one accommodation feature.
    \item Clear Connection: Explain how the feature addresses (or fails to address) the user's need, aligning with the prediction.
    \item Tone Consistency:
    \begin{itemize}
        \item Positive: Enthusiastic and encouraging, highlighting why it's a great match.
        \item Negative: Polite but clear, indicating why it's not suitable by pointing out mismatches or missing features.
        \item Neutral: Balanced and objective, noting mixed aspects or features that don't strongly align either way.
    \end{itemize}
\end{enumerate}
Strive for natural, helpful, and concise explanations."
\end{tcolorbox}

\begin{tcolorbox}[title=Amazon, colback=white, colframe=black, fonttitle=\bfseries]
\scriptsize
You are a Recommendation Specialist generating brief, natural-sounding explanations for movie or TV show recommendations. An AI model predicts user sentiment (positive, neutral, or dislike) based on how well the movie/TV show's attributes match user preferences.
Your task is to craft a concise justification for the model's prediction.

Input (JSON):
\begin{enumerate}
    \item prediction: The model's sentiment ("positive", "negative", "neutral").
    \item key\_features: Up to 5 key user preferences or movie/TV show attributes influencing the prediction (human-readable strings).
\end{enumerate}
    
Output (JSON): A single key, explanation, with a string value (max 20-30 words). You don't need to use all provided features.

Explanation Guidelines:
\begin{enumerate}
    \item User-Centric: Mention at least one user preference.
    \item Item-Specific: Refer to at least one attribute of the movie/TV show.
    \item Clear Connection: Explain how the attribute addresses (or fails to address) the user's preference, aligning with the prediction.
    \item Tone Consistency:
    \begin{itemize}
        \item Positive: Enthusiastic and encouraging, highlighting why it's a great match.
        \item Negative: Polite but clear, indicating why it's not suitable by pointing out mismatches or missing attributes.
        \item Neutral: Balanced and objective, noting mixed aspects or attributes that don't strongly align either way.
    \end{itemize}
\end{enumerate}
\end{tcolorbox}

\begin{tcolorbox}[title=Yelp, colback=white, colframe=black, fonttitle=\bfseries]
\scriptsize
You are a Recommendation Specialist generating brief, natural-sounding explanations for restaurant recommendations. An AI model predicts user sentiment (positive, neutral, or dislike) based on how well the restaurant's attributes match user preferences.Your task is to craft a concise justification for the model's prediction.

Input (JSON):
\begin{itemize}
    \item prediction: The model's sentiment ("positive", "negative", "neutral").
    \item key\_features: Up to 5 key user preferences or restaurant attributes influencing the prediction (these will be the string names of the features).
\end{itemize}
                
Output (JSON): A single key, explanation, with a string value (max 20-30 words). You don't need to use all provided features.
            
Explanation Guidelines:
\begin{enumerate}
    \item User-Centric: Mention at least one user preference.
    \item Item-Specific: Refer to at least one attribute of the restaurant.
    \item Clear Connection: Explain how the attribute addresses (or fails to address) the user's preference, aligning with the prediction.
    \item Tone Consistency:
    \begin{itemize}
        \item Positive: Enthusiastic and encouraging, highlighting why it's a great match.
        \item Negative: Polite but clear, indicating why it's not suitable by pointing out mismatches or missing attributes.
        \item Neutral: Balanced and objective, noting mixed aspects or attributes that don't strongly align either way.
    \end{itemize}
\end{enumerate}
Strive for natural, helpful, and concise explanations.
\end{tcolorbox}
\caption{Prompt templates used for generating FIRE’s explanations across all datasets.}
\label{fig:prompt-generation}
\end{figure}

To generate explanations within the FIRE framework, we first identify the most influential features contributing to the recommender’s decision and assign interpretable names to them. The explanation is then generated by conditioning the language model on these influential features, the predicted sentiment from the recommender, and the desired structural characteristics of a good explanation. The prompt templates used for this generation process are illustrated in Figure~\ref{fig:prompt-generation}.

\end{document}